\documentclass[preprint,3p,twocolumn]{elsarticle}

\usepackage{amsmath}
\usepackage{txfonts}
\usepackage[utf8x]{inputenc}
\usepackage[protrusion=true,expansion=true]{microtype}
\usepackage{hyperref,xcolor}
\usepackage[T1]{fontenc} 
\usepackage{graphicx}
\usepackage{ulem}
\usepackage{tikz}
\usepackage{enumitem}
\usepackage{color} 
\usepackage{array,multirow,makecell}

\graphicspath :
\graphicspath{{figures/}}

\begin{document}

\begin{frontmatter}

\title{Estimating the carbon footprint of the GRAND Project, \\a multi-decade astrophysics experiment }

\author[a]{Clarisse Aujoux}
\ead{clarisse.aujoux@gmail.com}
\author[a,b]{Kumiko Kotera}
\ead{kotera@iap.fr}

\author[c]{Odile Blanchard}
\ead{odile.blanchard@univ-grenoble-alpes.fr}

\author{\\for the GRAND Collaboration}

\address[a]{Sorbonne Universit\'{e}, UPMC Univ.  Paris 6 et CNRS, UMR 7095, Institut d'Astrophysique de Paris, 98 bis bd Arago, 75014 Paris, France}

\address[b]{{Vrije Universiteit Brussel, Physics Department, Pleinlaan 2, 1050 Brussels, Belgium}}

\address[c]{Université Grenoble Alpes, GAEL Laboratory (UMR CNRS 5313, UMR INRAe 1215), CS 40700 - 38058 Grenoble CEDEX 9, France}

\date{Received $\langle$date$\rangle$ / Accepted $\langle$date$\rangle$}

\begin{abstract}
 We present a pioneering estimate of the global yearly greenhouse gas emissions of a large-scale Astrophysics experiment over several decades: the Giant Array for Neutrino Detection (GRAND). The project aims at detecting ultra-high energy neutrinos with a 200,000 radio antenna array over 200,000\,km$^2$ as of the 2030s. With a fully transparent methodology based on open source data, we calculate the emissions related to three unavoidable sources: travel, digital technologies and hardware equipment. We find that these emission sources have a different impact depending on the stages of the experiment. Digital technologies and travel prevail for the small-scale prototyping phase (GRANDProto300), whereas hardware equipment (material production and transportation) and data transfer/storage largely outweigh the other emission sources in the large-scale phase (GRAND200k). In the mid-scale phase (GRAND10k), the three sources contribute equally. This study highlights the considerable carbon footprint of a large-scale astrophysics experiment, but also shows that there is room for improvement. We discuss various lines of actions that could be implemented. The GRAND project being still in its prototyping stage, our results provide guidance to the future collaborative practices and instrumental design in order to reduce its carbon footprint.
\end{abstract}


\end{frontmatter}


\section{Introduction}

Global warming poses critical risks to humanity and demands immediate actions in order to be mitigated. Anthropogenic greenhouse gas emissions, largely driven by economic and population growth, are the leading factors of the observed climate change. Reflections towards a drastic reduction of these emissions should be central to the conception of any scientific project.

In the near future, the climate crisis will certainly challenge the way astrophysicists carry out their research. Several working groups have been devising on solutions and their implementation at different levels in the international astrophysics community (e.g., \cite{williamson2019embedding,matzner2019astronomy,stevens2019imperative,barret2020estimating,Labo1.5}). This is however a new path, and efforts are to be carried out in terms of methodology, creativity, and experimentation.

Large-scale experimental projects are part of the building blocks of the astrophysics community. They gather a large fraction of the scientific staff and absorb a significant volume of the science budget. As such, it seems essential to assess their environmental impact. Besides, these experiments could turn out to be interesting for other laboratories to elaborate and test ideas, and to appreciate the best practices to be implemented in other contexts. 

We present here a rough estimate of the carbon footprint of the Giant Radio Array for Neutrino Detection (GRAND) project. GRAND is a planned large-scale observatory of ultra-high-energy cosmic particles — cosmic rays, gamma rays, and neutrinos with energies exceeding $10^8\,$GeV \cite{GRAND20}. The design of GRAND will be modular, consisting of several independent sub-arrays, each of 10\,000 radio antennas deployed over 10\,000 km$^2$ in radio-quiet locations. The collaboration consists today of 67 international researchers and engineers.

\begin{figure*}[!tb]
    \centering
    \includegraphics[width=0.95\linewidth]{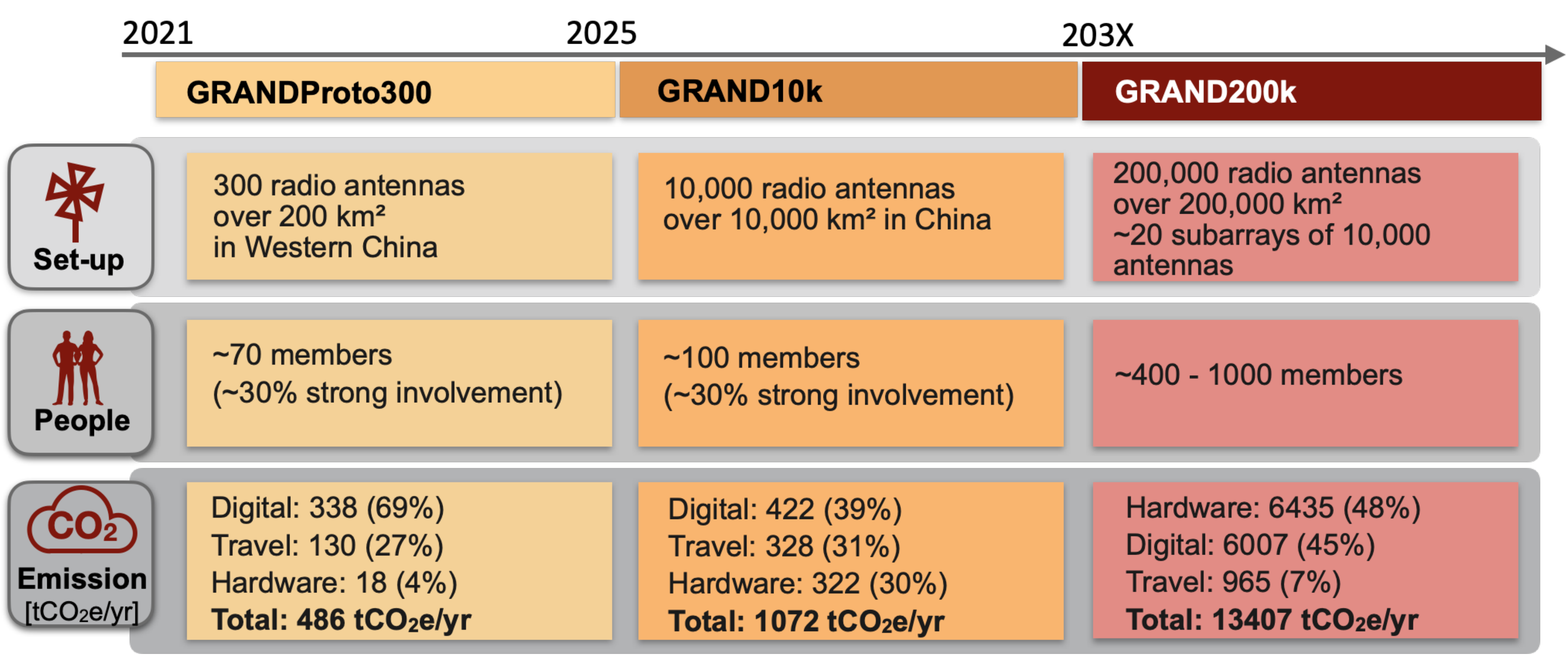}
    \caption{Roadmap of the GRAND project. The different stages of the project are presented, with information on the envisionned set-up, growth of the collaboration, and major greenhouse gas emission sources with their contribution in tCO$_2$e/yr and their corresponding percentage, as estimated in this work. }
    \label{fig:roadmap}
\end{figure*}

Within the collaboration, questions about the environmental impact of the project have arisen, and this is the object of this article. This paper focuses on the methodology used to assess the greenhouse gas (GHG) emissions linked to the GRAND project, and discusses the results and possible lines of actions, taking into account technological and scientific possibilities, and budgetary scenarios. 

This study is a pioneering attempt to assess the carbon footprint of a large-scale physics experiment. To our knowledge, no other global GHG assessment of a Physics experiment has been published so far, and the estimates given in this paper can serve as a basis for comparison and reflection. The specificity of this methodology is that it is fully transparent and uses open source data. Hence, the method is replicable to any other scientific consortium. The estimates given in this work are subject to large uncertainties related to emission factors and assumptions on future developments. The scope of this study is not to calculate precise carbon emissions, but to give estimates of the relative weight of each emitting category, in order to orient the action plans of the collaboration. 

Greenhouse gas assessments may cover numerous sources of emissions \cite{GHGprotocol}. In this study, three sources of emissions are examined: travel, digital technologies, and the hardware equipment. Those three categories were chosen because of their obvious importance in terms of emissions, but also because of the lever for action the collaboration has on them. Travel cannot be neglected, as one obvious source of emissions of the collaboration. The digital contribution is often underestimated, but is at the heart of the day-to-day scientific work performed by the bulk of the collaboration. The hardware equipment will be dominated by the numerous radio detection units, each composed of a radio antenna, its mechanical, electronics and power devices. The evaluation of their carbon footprint will be crucial to this study, as the antennas will be deployed in great numbers and over large areas. 

Note that these main factors (travel, digital, hardware) are not inherent to this project, but common to many astrophysics and particle physics projects.

According to the Intergovernmental Panel on Climate Change, in order to limit warming to 1.5°C, global net anthropogenic CO$_2$ emissions have to decline by about 45\% from 2010 levels by 2030, reaching net zero emissions  by 2050 \cite{IPCC18}. Interestingly, these timescales correspond to the GRAND roadmap, as the full-size experiment is expected to be deployed by the 2030s. By that time, it is possible that unforeseen technological progress will have happened in various domains that could be implemented in the experiment, such as travel, solar panels, the carbon footprint of electronic devices and data centers. Also by the 2030s, it is highly likely that taking into account the carbon emissions will be mandatory in all scientific projects. Today, the GRAND project is starting its prototyping phase, hence it is an ideal time to adapt the experiment to climate requirements. For all these reasons, this study appears timely and necessary. \\

We first present the GRAND projet and the collaboration in Section~\ref{section:GRAND} . We then introduce the general methodology of this study (Section~\ref{section:methodology}). We examine the three main sources of GHG emissions of the experiment, namely travel (Section~\ref{section:transportation}), digital technologies (Section~\ref{section:digital}) and hardware equipment (Section~\ref{section:hardware}). We work on projections for the three stages of the project (GRANDProto300, GRAND10k, GRAND200k) in a {\it Business as usual} scenario in Section~\ref{section:projections}. We propose and discuss action plans which could help to reduce the emissions compared to this scenario in Section~\ref{section:action_plans}, and give our conclusions and perspectives in Section~\ref{section:conclusion}.

\section{The GRAND Project and Collaboration}\label{section:GRAND}

\begin{figure*}[!tb]
    \centering
    \includegraphics[width=0.49\textwidth]{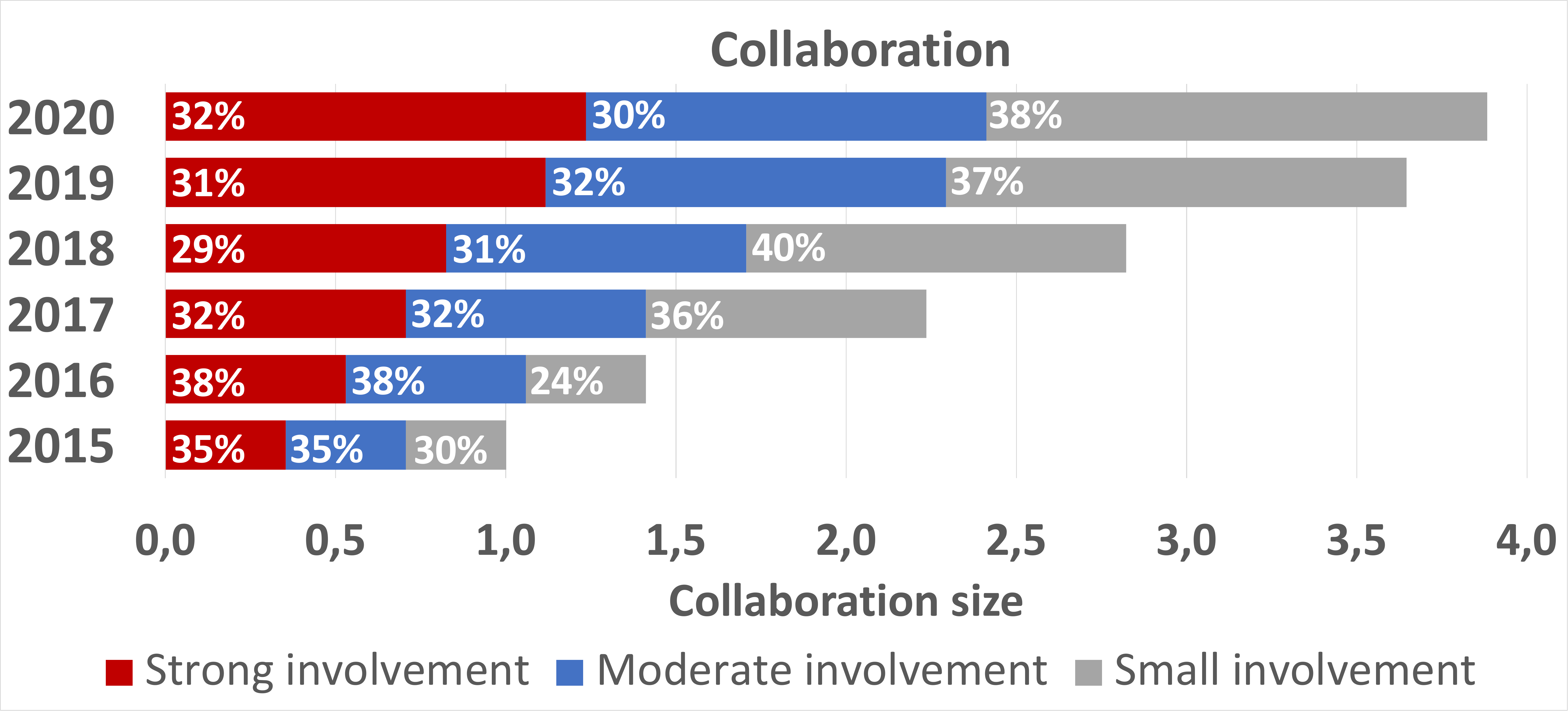}
    \includegraphics[width=0.49\textwidth]{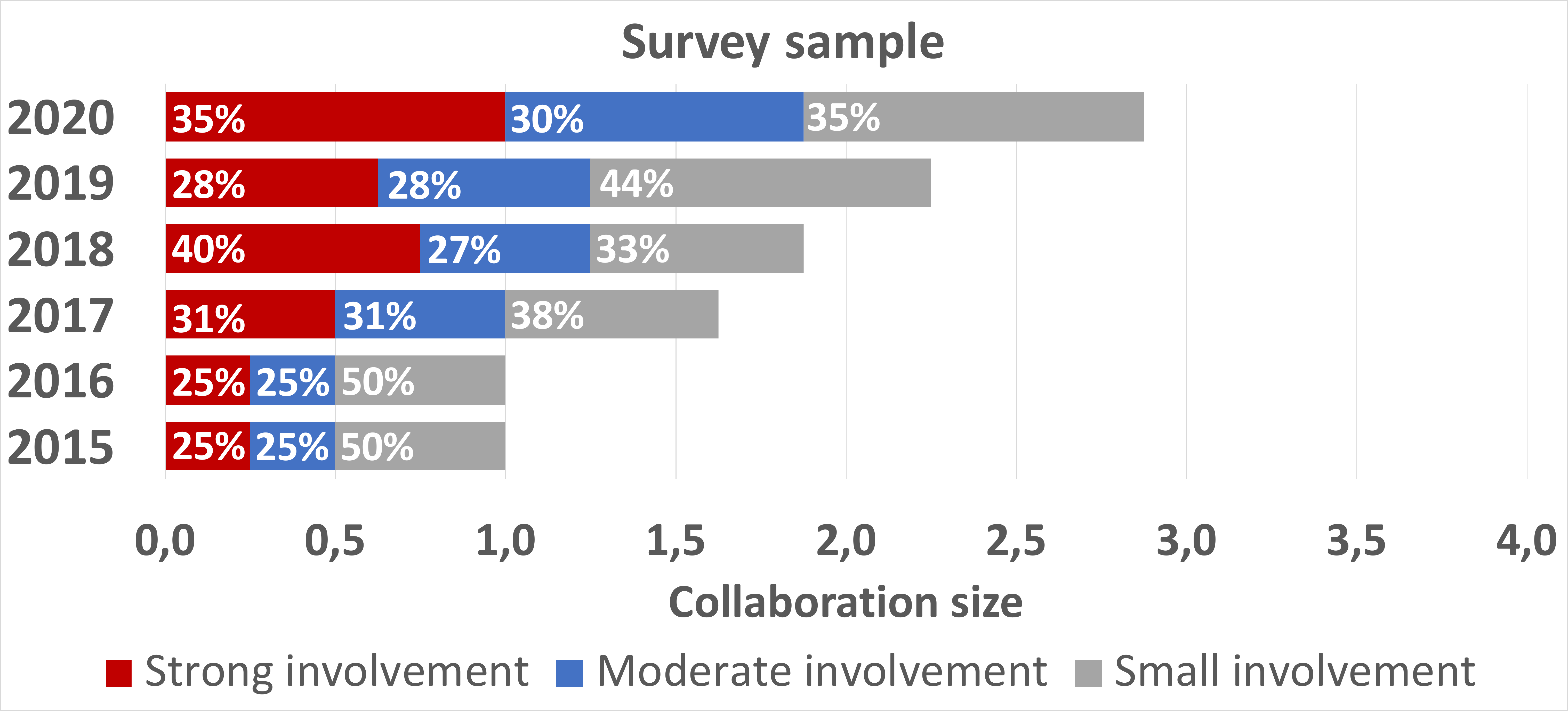}

    \caption{Trends  of the collaboration size over the years and of the fraction of members with strong ($>30\%$ of the working time), moderate ($10\%-30\%$) and light ($<10\%$) involvement in the project. The size of the collaboration is normalized to the initial number of members in 2015. The left panel analyzes the whole collaboration (67 members), and the right panel is based on a 33\% sub-sample, collected via a survey sent to the collaboration.}
    \label{fig:members}
\end{figure*}

In this section, we  first present some elements of the GRAND project and collaboration that are relevant to estimate the present carbon emissions, to estimate projections and discuss action plans.

\subsection{Project roadmap}\label{section:roadmap}

The GRAND project was initiated in 2014 by a handful of researchers. In order to discuss its feasibility and to attract interest, two workshops were organized in 2015 in Paris and in Chicago, from which a core of active scientists emerged. Since then, the collaboration has steadily grown to reach its current size of 67 members (see Section~\ref{section:collaboration}). 

The GRAND project will follow a staged construction plan, to ensure that key techniques are progressively validated. The collaboration is currently working on GRANDProto300, a 300-antenna prototype, which will be a test bench for the further steps of the experiment. The array is to be deployed in 2021 over 200\,km$^2$, in a desertic area in Western China. The following stage corresponds to GRAND10k, with 10\,000 antennas to be deployed around 2025. GRAND10k is designed to reach an ultra-high-energy neutrino sensitivity comparable to that of other potential contemporary detectors. The final stage, GRAND200k, will replicate GRAND10k arrays in different locations worldwide, in order to reach the ultimate target sensitivity of GRAND in the 2030s. 

The various stages of the project are depicted in Fig.~\ref{fig:roadmap}. The collaboration will likely grow to reach about 100 members for the GRAND10k stage. Beyond this stage, the collaboration is foreseen to reach $400-1000\,$members. This number is based on a sheer budget necessity (each member contributes to funding the running costs of the experiment), as it is the case in most astroparticle physics experiments (e.g., the Pierre Auger Observatory, the IceCube experiment). Besides a large number of contributors help achieving a diverse science case and obtaining quality physics results. 
GRAND collaboration meetings are organized every year to discuss the results and future plans. These annual meetings are held in a city harboring one of the GRAND-member institutes, or near potential experimental sites in China.

\subsection{Collaboration members}\label{section:collaboration}
The GHG emissions generated by the collaboration depend on the number of people involved in the project and the degree of their involvement. The level of involvement of a member can be evaluated as the percentage of the working time spent on GRAND-related work. Note that this level can also vary in time. 

In our framework, the knowledge of these levels for all members is a prerequisite to estimate the fraction of basic daily emission factors that should be attributed to the GRAND collaboration. Specifically, in Sections~\ref{section:transportation} and~\ref{section:digital} we calculate the share of emissions due to commuting and to the daily use of electronic devices (e.g., laptop, smartphone) associated to the project. 

Three populations can be identified in a collaboration. The distribution of these populations and their evolution will be essential tools for projections. The populations are defined according to the level of involvement of the members, i.e., their percentage of working time spent on GRAND-related tasks:
\begin{enumerate}[label=\Roman{*}.,noitemsep]
    \item Strong involvement: $>30\%$ of working time, 
    \item Moderate involvement: 10\% to 30\% of working time,
    \item Light involvement: $<10\%$ of working time.
\end{enumerate}
Members with a strong involvement are likely to massively travel for the GRAND project. The population consists of experimentalists traveling on-site for measurements and deployment, core-team members of the project who travel for administrative reasons and promote the project, active members, including fully involved students and postdoctoral scholars, who visit one-another to discuss and obtain results that they present at conferences. These members are also likely to physically join most annual collaboration meetings. 

Members with a moderate involvement will join some annual collaboration meetings, and occasionally participate in conferences to present GRAND-related results. 

Finally, members with a light involvement will rarely join the annual collaboration meetings physically, and will rarely travel for the primary purpose of communicating for GRAND.\\

The left panel of figure~\ref{fig:members} presents the evolution of the collaboration size over the years and the 3 populations listed above. These numbers were derived from the history of the collaboration memberships and by sorting the different members by involvement category. They enable us to evaluate any bias that a sub-sample of the collaboration might have compared to the full population, as discussed in Section~\ref{section:survey}.

One can see in Figure~\ref{fig:members} that the collaboration size has grown regularly since 2015, and has almost quadrupled in size by 2020. The proportion of the different populations has remained almost constant over time. 

As an international collaboration, GRAND members  originate from institutes located in several countries. The main countries presently involved are (in alphabetical order): Brazil, China, France, Germany, the Netherlands, and the United States. This geographical spread, not specific to GRAND but to any international collaboration, raises obvious concerns about communication (e.g., physically gathering collaborators regularly, and hence about travel, but also about the digital infrastructure).

\section{General methodology}\label{section:methodology}

GHG emissions are determined  by multiplying GHG-generating activity data (e.g., kilometers  traveled, electricity consumption, Central Processing Unit -CPU- time...) with  emission factors (e.g., CO$_{2}$e\footnote{Short for "Carbon Dioxide equivalent". There are several greenhouse gases other than CO$_2$ in the atmosphere: water vapour, methane (CH$_4$), nitrous oxide (N$_2$O). Each greenhouse gas has a different global warming potential (GWP), which indicates the amount of warming this gas causes over a given period of time. The GWPs used here are relate to a period of 100 years and refer to the IPCC 5th report values \cite{IPCC5}. For any quantity and type of greenhouse gas, CO$_2$e corresponds to the amount of CO$_2$ which would have the equivalent global warming impact \cite{CO2eq}.} emissions per kilometer traveled by plane, CO$_{2}$e emissions per kWh of  electricity consumed, CO$_{2}$e emissions per CPU hour...). This enables to aggregate all GHG emissions in one single indicator, called the "carbon footprint".

Due to various types of uncertainties, emission factors are only orders-of-magnitude estimates. Therefore, the results presented in this report are  given within orders of magnitude. All hypotheses and emission factors are mentioned, in order to allow for comparisons with other greenhouse gas assessments. 

As in any greenhouse gas assessment, uncertainties can be high depending on the origins of the data. We adopt the uncertainties of the Methodological Guide of the French Association Bilan Carbone \cite{methodo}. The uncertainties are as follows: 
\begin{itemize}[noitemsep]
    \item 0\% to 5\% for an emission factor from a direct measurement,
    \item 15\% for a reliable non-measured data,
    \item 30\% for a calculated data (extrapolation),
    \item 50\% for an approximate data (statistical data),
    \item 80\% for an order-of-magnitude estimate.
\end{itemize}

The emissions will be computed in units of CO$_2$e.\\

\section{Travel}\label{section:transportation}
 
The travel sector produced 8.2 Gt\,CO$_2$e of direct greenhouse gas emissions in the world in 2018 (including non-CO$_2$ gases)  and hence was responsible for approximately 24\% of total energy-related CO$_2$ emissions \cite{IEA}. 
With daily commutes and regular academic travels, researchers contribute to this source of emission. The impact of the daily commutes of the GRAND members has been computed along with the impact of academic travel. Although it is likely that the GHG emissions related to travel will be dominated by academic travel, it is interesting to be aware of their distribution between the two sources. This can help us to choose the actions that will have the greatest impact on the greenhouse gas emissions. 

GRAND scientists travel regularly for purposes that can be split into three main categories: 
\begin{itemize}[noitemsep]
\item Annual collaboration meetings, 
\item On-site missions, 
\item Other (conferences, visits, seminars, etc.). 
\end{itemize}

Indeed, as an international collaboration, GRAND members gather at yearly meetings across the world, in order to assess the progress and discuss the next steps of the project. It is also necessary for experimentalists to travel to China in order to make field measurements and to deploy the antennas. Scientists also travel to visit each other and hold specific face-to-face discussions, to promote the project and its results at conferences or at seminars.

\begin{figure}[!tb]
    \centering
    \includegraphics[width=1\columnwidth]{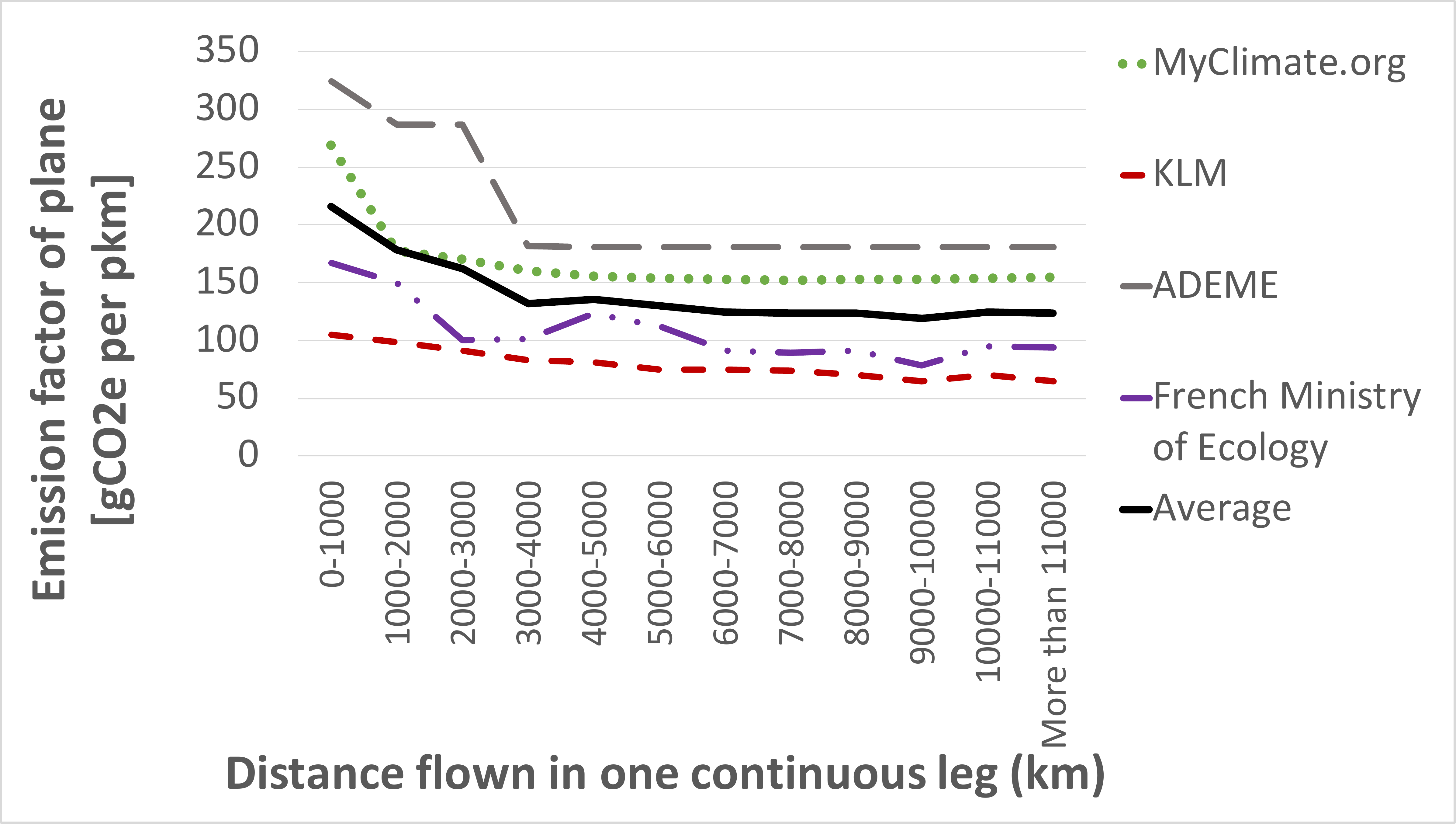}
    \caption{Different greenhouse gas emission factors for flights, depending on the database chosen. The flight distance is flown in one continuous leg. The black solid line corresponds to the average of the 4 datasets.}
    \label{fig:plane}
\end{figure}

\subsection{Survey and representativeness}\label{section:survey}

A survey was sent to the whole collaboration with the primary purpose of assessing the carbon emissions of  travel within GRAND. The results are fully studied in this section and also used in Section~\ref{section:digital}. 

The survey aims at evaluating the commuting and travel frequency and distances of the members.
Each member was asked to provide a list of all the trips undertaken for the project, mentioning their purpose (annual meeting, on-site mission or else), and the means of travel. As for commuting, the number of kilometers travelled for commuting and the means of travel were asked to the collaborators.

In what follows, we discuss the representativeness of the results. 
The survey was sent to all 67 members of the GRAND collaboration, and 33\% responded. The members were asked to evaluate their level of involvement in the collaboration since 2015. This information enabled us to categorize the respondents within the three involvement populations described in Section~\ref{section:collaboration}. The right panel of Figure~\ref{fig:members} presents the evolution of the collaboration size and the three populations over $2015-2020$, as estimated from the survey respondents.

Given the small sample of answers received, it is difficult to use inferential statistics to extend the results to the whole studied population. As a first approximation, one can assume that the respondents have similar characteristics on average as the global population. An overall correction factor of $\sim 3$ can then be applied to global estimates obtained using the survey, in order to correct for the limited sample size, given that the proportions of the involvement populations are roughly well-represented. 

In detail, we note that the growth of the collaboration size is underestimated in the survey by a factor of $3/4$. In particular, the growth between the years 2015 and 2016 are not represented in the survey. However, this correction is sufficient for our purposes, as we are interested in orders-of-magnitude estimates.

\subsection{Travel emission factors}

The data used for the emission factors is divided into two categories : air travel and other travel modes. For the other travel modes (car, public transportation, etc.), we use  the ADEME carbon database~\cite{ADEMECarbonDatabase}. This database has several advantages: it covers a wide range of emission factors and it is easily accessible. However, this is a database centered on French specificities. For example, for public transportation such as the subway, the fuel mix for electricity generation is specific to France which, unlike other countries, has a low carbon intensity due to the use of nuclear power (see~\ref{app:Emix}). Still, the difference between the carbon intensity of electricity in commuting in different countries does not affect the final result for two main reasons: commuting represents only a small fraction of overall travel emissions, which are dominated by air travel;  the main type of travel used for commuting by GRAND members is car (gasoline and diesel) which represent 45\% of the kilometers traveled. 

Concerning aviation, its effects on climate change are more than just emissions of CO$_{2}$ from burning fuels. It is the result of three types of processes: direct emissions of radiatively active substances, emissions of chemical products that produce or destroy radiatively active substances, and emissions of substances that trigger the generation of aerosol particles or lead to changes in natural clouds~\cite{IPCC1999}. Those effects are not always taken into account in carbon databases that include emission factors for planes, as they are not yet very well understood. 

Choosing one single database can lead to highly different results, up to a factor of 5 \cite{barret2020estimating}. To compute our travel carbon footprint, four databases were chosen from different organizations: the database from ADEME \cite{ADEMECarbonDatabase}, the database from the French Ministry of Ecology \cite{MinistryEcology}, data from KLM airline \cite{KLM} and a computation methodology from MyClimate.org \cite{MyClimate}. Those four databases have diverse sources: one non-governmental organization, one airline company, and two state organizations. 

In order to have more accurate results, the emission factors are split in categories depending on the number of kilometers travelled by increments of 1,000\,km up to over 11,000\,km. The data used is already divided in such categories for the French Ministry of Ecology and ADEME. MyClimate.org divides the emission factors between short-haul flight and long-haul flights. The KLM data is a list of all the flights they offer, their distance in kilometers and the greenhouse gas emission associated to the flight.

Figure~\ref{fig:plane} presents the range of flight emission factors spanned by the various datasets. The greenhouse gas emissions as a function of the distance flown in one continuous leg are presented for the 4 databases described above (ADEME\footnote{The emission factors from ADEME were retrieved from the database prior to the June 2020 update by ADEME.}, MyClimate.org, French Ministry of Ecology, KLM). The uncertainty of these emission factors is estimated at around 60 \% in the ADEME database and mainly stems from the uncertainty pertaining to the emissions of contrails. The other databases do not provide such estimates. The black solid line corresponds to the average value calculated from the 4 databases, which is used in our emission estimates for one-legged trips. We have not provided uncertainty estimates for this average value as only the ADEME database provides some. It is worth mentioning that the average emission factors are respectively between 5 and 10 \% greater when considering respectively two or three legs instead of one-legged long haul trip.

\subsection{Commuting}

In order to compute the emissions due to daily commutes, we first need to estimate the number of days worked by the GRAND collaboration members. 
For this purpose, questions were asked in the survey about the percentage of worked days dedicated to GRAND, the percentage of days worked remotely from home, the country of residence (which gives the number of working days per year in the country) and the number of years the collaborator has been working for GRAND.

The number of kilometers travelled for commuting for each type of travel was computed using the number of days worked per person annually and the number of kilometers travelled per person per day by mode of travel.

\begin{figure}[tb!]
    \centering
    \includegraphics[width=1\columnwidth]{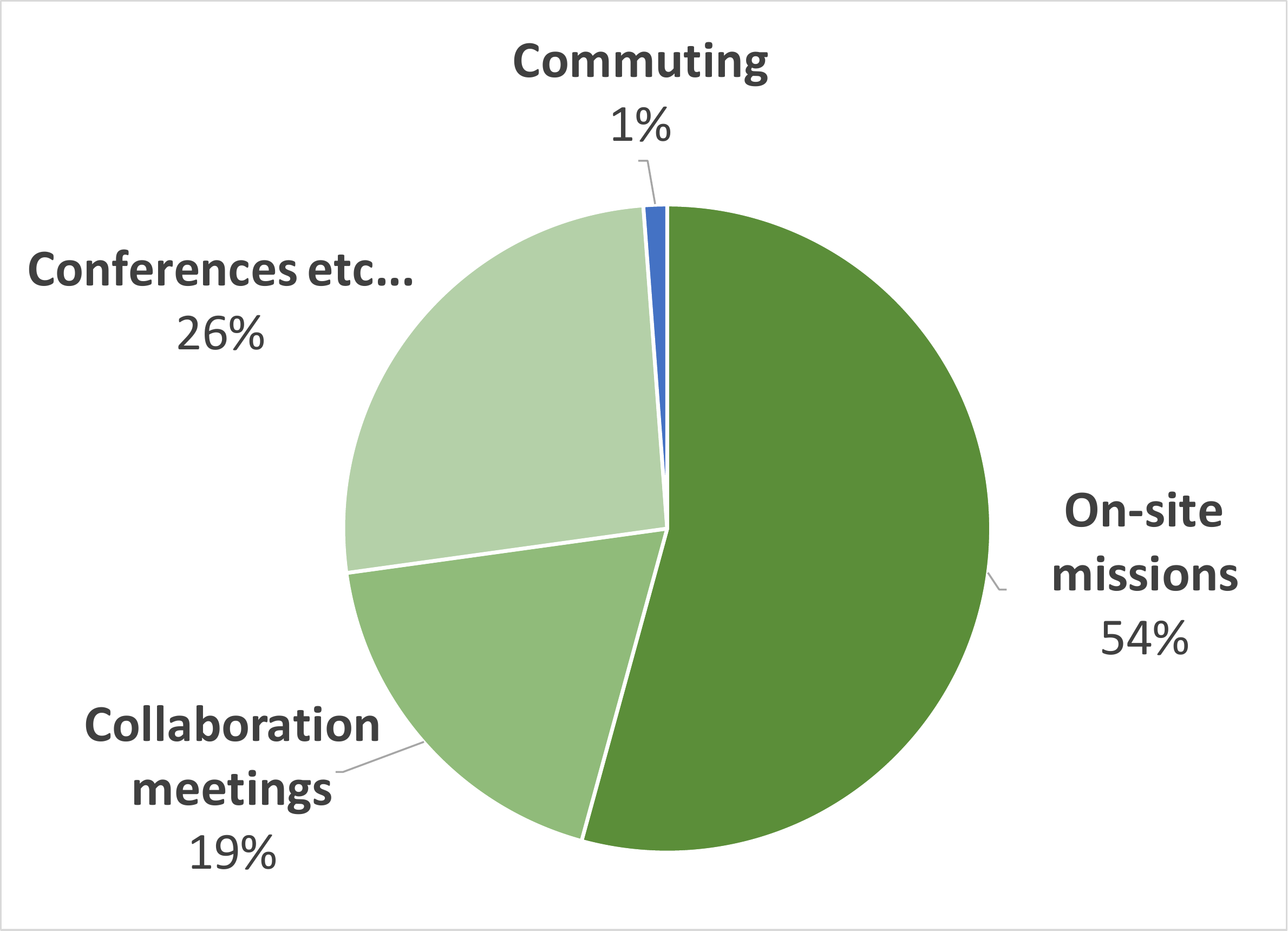}
    \caption{Distribution of emissions related to travel: on-site missions, collaboration meetings, conferences, commuting, from 2015 to 2020.}
        \label{fig:travel_emission_total}
\end{figure}

\subsection{Results}
\begin{figure}[tb!]
    \centering
    \includegraphics[width=1\columnwidth]{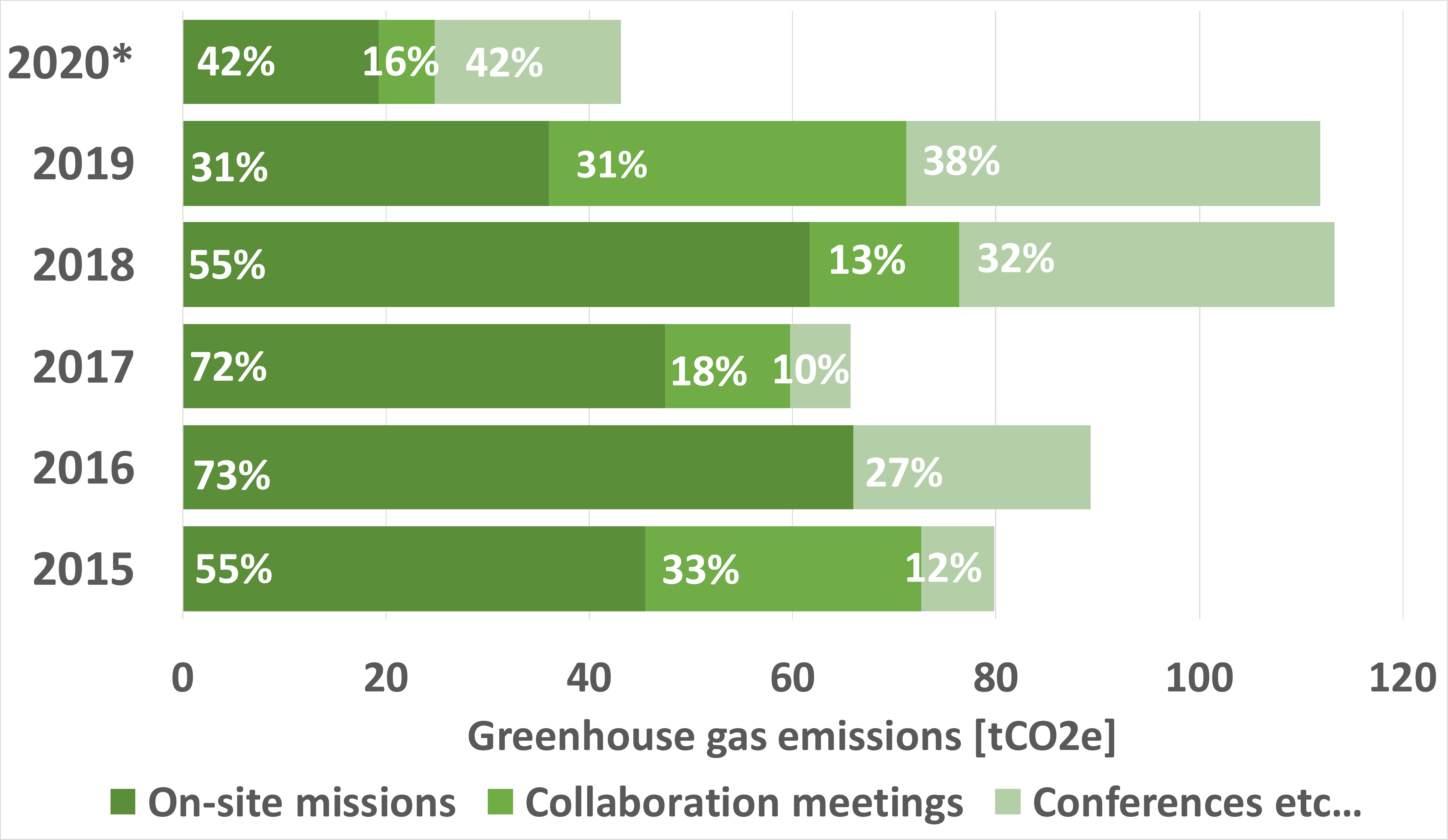}
    \caption{Trends of travel emissions over the years based on the survey responses corrected to an overall factor of $\sim 3$. The percentage attributed to three travel categories (on-site missions, collaboration meetings, other) is indicated on each histogram. *The emissions for 2020 are restricted to the first two months of 2020.
    }
    \label{fig:travel_emission_year}
\end{figure}

In this section we present the results stemming from our survey. We recall that 33\% of the collaborators answered the survey and the representativeness of the survey was discussed in Section~\ref{section:collaboration}. The results are summarized in Figs.~\ref{fig:travel_emission_total} and~\ref{fig:travel_emission_year}.

All in all, from 2015 to 2020, a total of 170 tCO$_2$e were emitted for all categories  in the survey sample. As discussed in section \ref{section:survey}, an overall  factor of 3 is applied to global estimates obtained using the survey, in order to account for the whole panel of collaborators. Using this multiplication factor, the total greenhouse gas emissions of GRAND for the $2015-2020$ period amounts to 510\,tCO$_{2}$e.

Most of the emissions stem from academic travel (99\%) and only 1\% are generated by commuting, as shown in Figure~\ref{fig:travel_emission_total}. Interestingly, among the travel categories, on-site missions constitute the dominant item. This seems to be consistent with a nascent experiment which requires to perform site surveys, as well as tests and hardware deployment in remote areas in China. As a benchmark figure, one Paris-Dunhuang flight emits approximatively 1.7 tCO$_2$e (Dunhuang, Gansu Province, China, being the closest airport to the experimental site of GRANDProto300).

Figure~\ref{fig:travel_emission_year} presents the trends of the travel emission sources over the years  for the survey sample. One can clearly see the experiment taking off in $2018-2019$, with maximal yearly levels of  $\sim 114$\,tCO$_2$e, due to the development of the prototype GRANDProto300, which was planned to be deployed in 2020. The emissions in 2020 pertain only to the months of January and February, after which the Covid-19 pandemic froze most academic travels, in particular to China. The proportion between the various categories of travel fluctuate as a function of the need for on-site missions (e.g., a large site survey campaign was conducted in $2017-2018$), and of the location of the annual collaboration meeting (e.g., the meeting of 2019 took place in Dunhuang, China, hence the increased emissions). 

\begin{figure}[tb!]
    \centering
    \includegraphics[width=1\columnwidth]{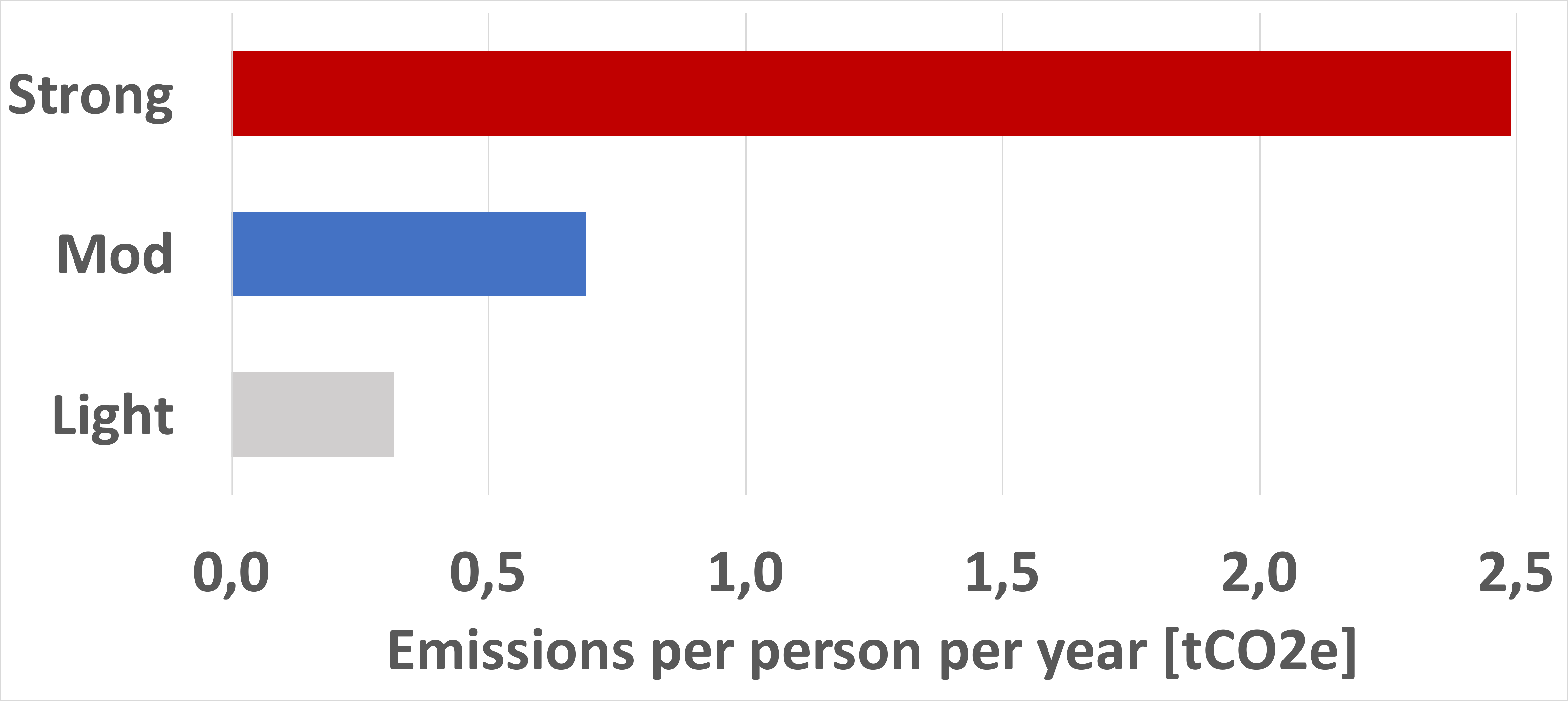}
    \caption{ Average travel emissions per year for the three types of GRAND collaboration members as described in Section~\ref{section:collaboration}: strongly involved, moderately involved, lightly involved.}
    \label{fig:emission_population}
\end{figure}

As discussed in section~\ref{section:collaboration}, the level of involvement in the collaboration results in different traveling profiles. This is illustrated in figure~\ref{fig:emission_population}, where the emission for three typical members stemming from the three involvement populations (light/moderate/strong) are presented. These results were estimated from the survey. Strongly involved members emit on average over 3 times more than moderately involved members. Theses figures, combined with a potential future distribution of the different populations, will enable us to derive emission projections for the next steps of the GRAND experiment.

\section{Digital technologies} \label{section:digital}

Digital technologies are often proposed as a solution for climate change mitigation: automation and connectivity (e.g., smart grids, connected mobility...) are presented as a way to reduce energy consumption. However, as reported for example by the Shift Project~\cite{ShiftProject}, digital technologies have a non-negligible environmental impact. Carbon footprint assessments often neglect the digital contribution, and the impact of the digital equipment is largely underestimated by its users. 

Digital technologies are used in every aspect of the GRAND project: e.g., for simulations, data analysis, data storage and transfer, and communication. It is hence necessary to evaluate the digital impact on greenhouse gas emissions, in order to compare it to the other emission sources. For such a study, defining the scope is essential. A focus has been made on four categories, which we present in the following sections: electronic devices, communication \& data exchanges, numerical simulations, and data transfer, processing and storage.

\subsection{Electronic devices}
\label{section:electronic_devices}
Among the electronic devices owned by GRAND members, laptop, desktop computers and screens are most prominently used for the GRAND project. A limited number of collaborators work on a digital tablet and the percentage of smartphone usage dedicated to the project is on average negligible compared to the laptop  and desktop computers.

To assess the carbon footprint of electronic devices, we use the online tool Ecodiag developed by the CNRS~\cite{ecodiag}. It allows us to compute the emissions related to various electronic devices such as laptop, desktop computers, etc., for given lifetimes, as presented in Table~\ref{tab:devices}. Each device has a carbon footprint corresponding to its production phase and its utilization phase. As the carbon footprint of keyboards, mouses, headphones and other accessories are negligible compared to the  contribution of computers, we do not take the former into account. The emission factor for the electricity consumption of the devices is the European one, retrieved from ~\cite{ADEMECarbonDatabase} \footnote {The fraction of European collaboration members prevails at the current stage of the GRAND experiment ($\gtrsim 50\%$), although the fraction of members from China (with a higher electricity emission factor) is steadily growing ($\sim 25\%$). The presence of several members from Brazil ($\sim 15\%$), a country with low electricity emission factor, also balances the overall factor.}.

As the devices are not used only for GRAND-related projects, the percentage of each member's total work days dedicated to GRAND is taken into account. For example, for a member working $x$\% of the time for GRAND on two personal computers, the GHG emissions associated to those devices will be computed as $x$\% of the total emissions associated to the devices.

\begin{figure}[tb!]
    \centering
    \includegraphics[width=1\columnwidth]{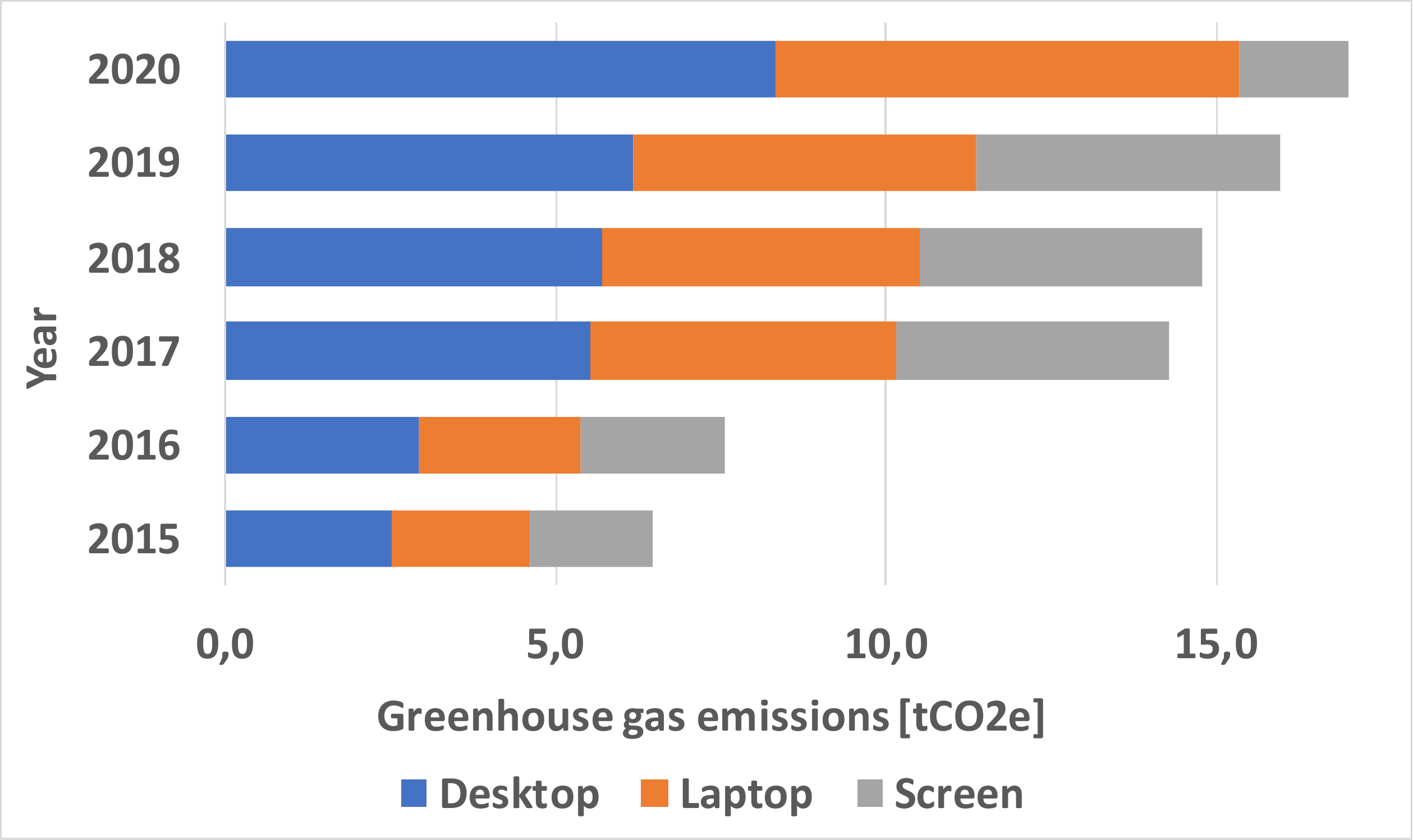}
    \caption{Cumulative greenhouse gas emissions related to the use of electronic devices by collaboration members, computed from the answers to the survey. An overall factor of $3$ is applied to the emission levels in order to correct for the survey-sampling bias.}
    \label{fig:devices}
\end{figure}

We assess emissions related to electronic devices based on the responses from the survey detailed in Section~\ref{section:survey} concerning the involvement level of each member. Our calculations assume that a collaboration member uses: one laptop (MacBook Pro 15'' --a large fraction of GRAND members are Apple users), one mid-range desktop and one screen\footnote{This can be viewed as a conservative assumption, as many members with Apple laptops do not use a desktop. However, more and more members use two screens (one at the office and one at home), and this is a growing trend with the incentive to work remotely, due to the Covid-19 situation. The emissions of a screen correspond roughly to that of a desktop, hence our calculation remains valid. A more precise investigation of the average computing devices owned by physicists would require a dedicated study.}. For the lifetime, we chose the shorter lifetimes listed in Table \ref{tab:devices}. Figure~\ref{fig:devices} presents the emissions related to the use of devices since the beginning of the collaboration.  These emissions correspond to the emissions of the whole collaboration, a factor of 3 was applied to the results of the survey. One can see that the emissions are consistently proportional to the growth of the collaboration. 

\begin{table}
\centering
\begin{tabular}{lrr}
  \hline
  Device & Lifetime & Emissions\\
  &  [years] & [kgCO2e/yr] \\
  \hline \hline
  \multirow{2}*{MacBook Pro 15''} & 3 & {130.2} \\
   & 4.5 & 93.2 \\
  \hline
  \multirow{2}*{Average desktop} & 4 & {154.4} \\
   & 6 & 129.4 \\
  \hline
  \multirow{2}*{Average screen} & 5 & {115.4} \\
   & 7.5 & 86.4 \\
  \hline
  
  \hline
\end{tabular}
\caption{Devices and their overall greenhouse gas emissions, including production and utilization phases.  The emission factor used  for the electricity consumption is the European one}. Emissions are listed for two typical lifetimes. Source: Ecodiag-CNRS~\cite{ecodiag}.
  \label{tab:devices}
\end{table}

These estimates are conservative, as we systematically choose the higher emission ranges for the devices used by each member of the collaboration. 

\subsection{Online exchanges}
Online exchanges can take different forms: email, video call, team communication platforms, cloud. These media do not have the same impact on carbon emissions, and have to be compared to other solutions, as, for example, gathering everyone in the same room. 

Assuming that each member of the collaboration sends 3 emails every working day during the time working for GRAND, it amounts up to 180\,000 emails sent since 2015. 
To estimate the number of days worked for GRAND per person, we assumed an average number of days worked annually per country, and for each member of the collaboration, we weighted this number with the percentage of time worked for GRAND.

The Shift Project estimated that sending an email with a 1\,MB attached file (a conservative hypothesis in our case) leads to an electricity consumption of 1\,Wh \cite{ShiftProject}. Using the  emission factor of the European electricity consumption ~\cite{ADEMECarbonDatabase}, as explained in section \ref{section:electronic_devices}, the total emissions amount to 74\,kgCO$_{2}$e for the whole period $2015-2020$. With a higher rate of 20 emails per working day sent by each collaboration member during the time working for GRAND, which is a conservative hypothesis, the emissions amount up to $\sim 0.5$\,tCO$_{2}$e for $2015-2020$.  This is still negligible relative to the other sources of emissions.

Monthly collaboration meetings, regular core-team meetings and other discussions are held via video-conferencing. The environmental cost of video-conferencing is highly dependent on several factors, such as the meeting duration, the technologies used, etc. The emissions associated to video conferencing are estimated to be 7\% of the ones from in-person meetings \cite{video}.  Refs.~\cite{2020NatAs...4..823B,2020Natur.583..356K} estimated differences in emission of several orders of magnitude ($\sim 3000$) between large-scale conferences held online or in person, in astronomy and in geophysics respectively. Although the scales of these conferences gathering several hundreds of participants are different to the smaller-scale ($\sim 30$ participants) meetings mentioned here, these estimates confirm that, compared to the other emission sources, the impact of the use of video-conferencing in the collaboration can be neglected. Note that this technology is also often presented as a greener option than in-person meetings, and its use could be considered for the GRAND annual collaboration meetings (see Section~\ref{section:proj_transportation}).

\subsection{Numerical simulations and data analysis}\label{section:simulations}

\begin{figure}[tb!]
    \centering
    \includegraphics[width=1\columnwidth]{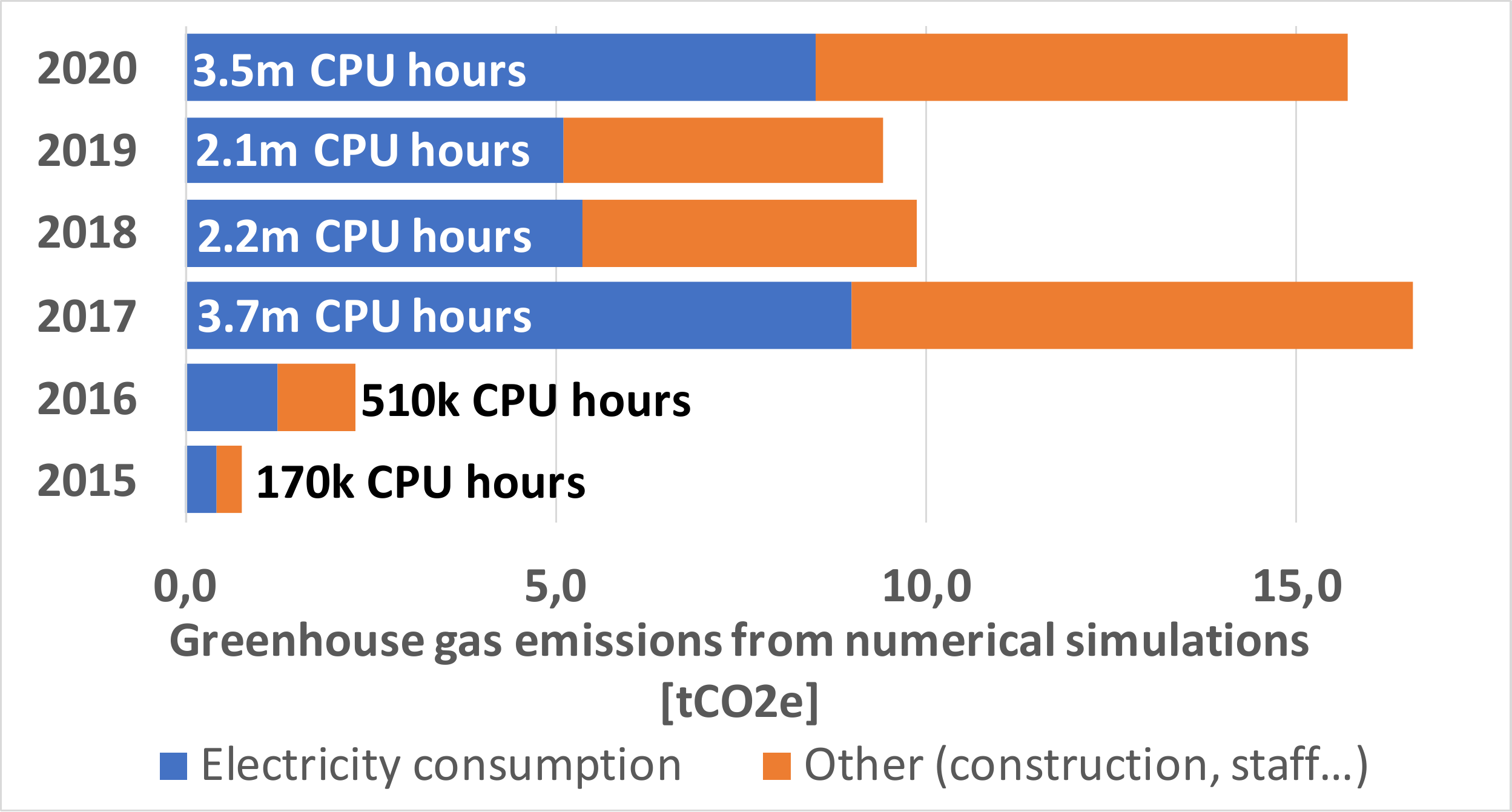}
    \caption{Emissions from numerical simulations run for GRAND. Emissions are separated in two categories in order to understand the impact of electricity consumption and other sources of emissions. }
    \label{fig:simulations}
\end{figure}

Numerical simulations are essential at all stages of the project, and particularly for the preparatory phases.
Over the last four years, about 2 million  CPU hours were used on various superclusters (principally at CCIN2P3 in Lyon, ForHLR at the Karlsruhe Institute of Technology, and UFRJ in Rio de Janeiro). Using the emission factors from Ref.~\cite{Berthoud}, we computed the emissions associated to simulations from 2015 to 2020, as presented in Fig.~\ref{fig:simulations}. The emission factors take into account all the stages related to performing simulations in a data center: server manufacturing, electricity consumption, and emissions generated by the staff running the data center. In our calculations, we used the electricity-specific emission factor of the countries hosting the superclusters where the simulations were run (France, Germany and Brazil, see~\ref{app:Emix}).

The scope chosen has a great impact on the results of the greenhouse gas assessment. In order to see the impact of every stage, we present the emissions related to electricity consumption only, and to the larger scope including staff interventions, server manufacturing and the like. 

The storage of the results of the simulations also have a cost in terms of GHG emissions. The percentage of the energy use in data center for data storage is around 10\% of the emissions of simulations \cite{USA_Storage}. We thus estimated the energy consumption related to the storage of the results of the simulations at around 10\% of the one related to simulations.

Once experimental data has been collected (see next sections), it will be analyzed. For GRANDProto300, each year over 5 years, between 200k CPU hours and 400k hours will be needed to analyze the astroparticle data. This amounts to 0.9\,tCO2e to 1.7\,tCO2e per year for GRANDProto300, using the same emission factors as for simulations. Data analysis is comparatively one order of magnitude less emitting than the simulations run during the preparatory phases. 

Here we did not take into account the analysis required to work on the Radio Astronomy science case, and focussed on the primary science case of GRAND, which is the detection of ultra-high energy astroparticles.

\subsection{Data transfer}\label{section:data_transfer}

 In this study, we focus on the data transfer related to the first prototype of GRAND, GRANDProto300, which is being currently developed. The projections for the next phases are discussed in Section~\ref{section:proj_digital}. 

On the GRANDProto300 experimental site, data will be collected by the radio antennas deployed . The data collected will be sent to the nearest city via Ubiquiti AirFiber \cite{ubiquiti}. The raw data will amount roughly to 10\,TB per day. Assuming that the data acquisition will take place over 300\,days a year (due to weather conditions, maintenance, etc.) during 5 years, about $15\times 10^3$\,TB of data will be collected during the whole experiment. We calculate the GHG emissions that would be produced if this data were to be transferred via the Internet to a data center in Beijing and then transferred across the world to a few data centers harbored by GRAND institutes.

The computation of the emissions is based on the electricity intensity of Internet data transmission which represents the amount of electricity consumed per amount of data transmitted. This quantity can vary much according to the way it is computed: the system boundaries, the assumptions made, and the year to which the data apply, significantly affect the estimates. Reference~\cite{InternetAslan} provides an estimate of 0.06\,kWh/GB for 2015. Using this number along with the estimation of the quantity of data transferred as given above, we evaluate the total consumption of data transmission to about$\sim900$\,MWh.

The emissions related to this energy consumption depend on the emission factor  chosen for electricity consumption. For example, the European electricity emission factor is 0.42 kgCO$_{2}$e/kWh while the Chinese one  is 0.766 kgCO$_2$e/kWh \cite{ADEMECarbonDatabase}, due to different energy mixes to generate electricity. In order to choose an emission factor as close as possible to reality, we take into account the distribution of data center locations across the world, as the data will travel around the globe to reach the various GRAND institutes. We assume  that the data centers are located in the same country as the data transmission network equipment. The countries hosting the largest number of data centers are the United States (38\% of the centers), along with five other countries which host together 55\% of the data centers in the world (see ~\ref{app:data_centers}). The emission factor is calculated as the weighted average of the emission factors of these countries, based on the proportion of data centers hosted. Note that this calculation does not account for the size of the data centers, which can vary from one country to another. The emission factor obtained yields an emission of 0,518\,kgCO$_{2}$e/kWh.

Finally, we calculate that the total electricity consumption needed for the raw data transmission for GRANDProto300 emits 470\,tCO$_{2}$e. This amount assumes that the data collected is only transferred once.  If we assume that 10 institutes will transfer the data from the central data center in Beijing, emissions amount to 4700\,tCO$_2$e. \\

Interestingly, for the primary scientific scope of the experiment, i.e., radio-detection of astroparticles, the relevant data could be efficiently reduced on-site down to 100\,GB per day, leading to 150\,TB of data transmitted during the whole experiment. The transmission of this limited volume of data would imply 2 orders of magnitude less emissions, leading to a total of 4.7\,tCO$_2$e emitted via data transfer at the GRANDProto300 stage. The choice of keeping only this limited data reduces however the science case of the GRAND experiment. For the prototyping stage, it also appears cautious to keep all the available data. In Section~\ref{section:proj_data}, we give strategies to reduce the overall data amount for the later phases of the experiment.

\subsection{Data storage}\label{section:data_storage}

The data collected will be stored in cloud servers. Cloud servers allow all the collaborators to easily have access to the data and to process them with virtual machines. The collaboration envisions that the data be stored at 3 different locations to provide back-up. The peak consumption of cloud storage is around 11.3W/TB of data \cite{cloud_storage}. For one TB of data for a year, this leads to 99\,kWh of electricity consumption. Assuming the same emission factor of 0.518\,kgCO$_2$e/kWh as above, it leads to an estimate of 51.2\,kgCO2e for one TB of data stored for a year. Each year, GRANDProto300 stores 3,000 TB of data, which leads to around 153.6\,tCO2e/yr for GRANDProto300.

\section{Hardware equipment}\label{section:hardware}

In the GRAND project, the hardware equipment will be dominated by the radio detection units. The units will be deployed in large numbers (200\,000), hence their environmental impact cannot be neglected. A complete life cycle assessment of the hardware equipment is out of the scope of this work, and will be evaluated in a subsequent study. Still a rough assessment of the carbon footprint of the detection units can already initiate discussions about the impact of the hardware on the carbon footprint of the collaboration.

\begin{figure}[tb!]
    \centering
    \includegraphics[width=1\columnwidth]{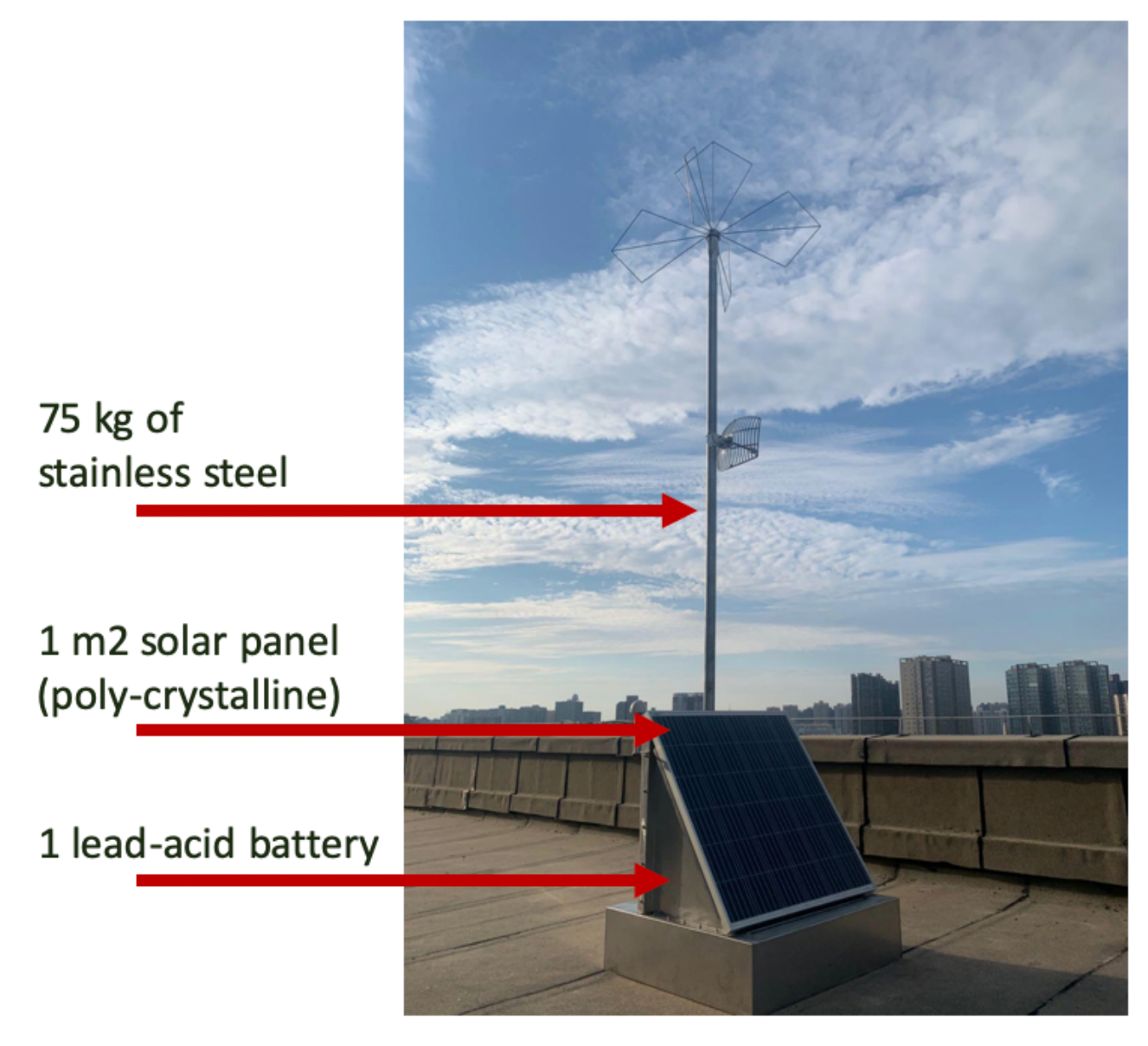}
    \caption{GRANDProto300 prototype antenna.}
    \label{fig:antenna}
\end{figure}

Each detection unit consists in a radio antenna with a mechanical support setup (made primarily of 75\,kg of stainless steel), data acquisition electronics (the material of the Print Circuit Board is MEGTRON \cite{PCB}), several cables of 6-meter length in total, a solar panel (poly-cristalline module) and a battery (see Fig.~\ref{fig:antenna}). 

Here we focus on three major parts of the detection units: the antenna with its mechanical setup, the solar panel and the battery.  We also focus on the hardware already designed for the prototyping phase GRANDProto300. The projections for the further stages of the experiment are presented in Section~\ref{section:proj_hardware}.

Note that for each of these items, recycling has not been taken into account in the GHG emissions. A recycling plan has not been elaborated yet within the GRAND collaboration, although it will clearly be an important line of action (see Section~\ref{section:proj_hardware}).

\subsection{Antennas}
In order to evaluate the carbon footprint of the antennas, we focused on the 75\,kg of stainless steel composing the arms, the frame, and the pole. The emission factor of stainless steel varies from $\sim 2.2$\,kgCO$_{2}$e/kg \cite{ADEMECarbonDatabase} to $6.8$\,kgCO$_{2}$e/kg \cite{metal} in the literature, depending on the assumptions on the production methods and the country. References before 2010 tend to indicate higher emission factors (see also Ref.~\cite{JOHNSON2008181}). Recent estimates of stainless steel production in China (where the first 10'000 units will likely be produced) give an emission factor as low as $1.76\,$\,kgCO$_{2}$e/kg. We choose to take a mid-range value of 2.9\,kgCO$_{2}$e/kg for this emission factor, closer to recent estimates, that correspond for example to the estimates documented in Ref.~\cite{ISSF2019}.

One antenna has therefore a carbon footprint of 217\,kgCO$_{2}$e. For the 300 units of GRANDProto300, this yields 65\,tCO$_{2}$e, and for the last stage of the project 43,400\,tCO$_{2}$e. 

It seems important to remind here that even if this figure appears to be high, it should be compared to similar infrastructures. Indeed, compared to other metals, stainless steel has a low emission factor. For example, aluminum has an emission factor of 35.7 kgCO$_{2}$e/kg ~\cite{metal}. However, the environmental impact of stainless steel depends on many external factors, that can lead to changes in its environmental impact. For example, the ore grade, electricity energy source, fuel types, and material transport as well as process technology influence the environmental impact of the extraction of this metal~\cite{metal}. These factors can vary in time, and by 2035 the emission factors used here to compute the emissions for GRANDProto300 could be completely different for GRAND200k.

\subsection{Solar panels}
Each antenna is fueled by a 1\,m$^{2}$ poly-crystalline solar panel. GHG emissions from photovoltaics show high variations even for photovoltaic systems from the same year. The current GHG footprint is estimated at around $\sim 20$\,gCO$_{2}$e per kWh for 1\,m$^{2}$ poly-Si photovoltaic systems ~\cite{pv}. The values used here refer to standardised conditions. The yield of a solar panel is usually 13\%, which leads to an annual production of 221 kWh per 1\,m$^{2}$. Using the emission factor presented above, one 1\,m$^{2}$ solar panel emits 4.42\,kgCO$_{2}$e per year. For GRANDProto300, this represents 1.3\,tCO$_{2}$e per year. However, a clear decrease in the environmental footprint of photovoltaic systems over time has been observed: the GHG footprint of photovoltaic systems were divided by 7 over 24 years~\cite{pv}. Following this trend, it is likely that  the GHG footprint of solar panels will decrease over the next decades.

\subsection{Batteries}
The batteries chosen for GRANDProto300 are lead-acid batteries. Here we only focus on their GHG emissions, but it should be pointed out that the environmental cost of lead-acid batteries should also be taken into account in terms of metal depletion and fossil fuel depletion\footnote{The environmental impact of lead-acid batteries is negative in case of grid-connected systems \cite{Lead-acid},  but such a setup is inapplicable to our study, as our experiment is carried out in a remote area without electricity supply.}

In terms of GHG emissions the impact of the production of lead-acid batteries per kilogram of battery   is 0.9\,kg\,CO$_2$e \cite{Lead-acid}. The batteries used for GRANDProto300 weigh 51\,kg, which yields an emission of 13.7\,tCO$_{2}$e for the whole array.

As already mentioned, no recycling plan has been elaborated yet by the collaboration, but this will be carefully discussed in the future. Indeed, the recycling of lead-acid batteries in China is a great concern for public health, as lead is classified as one of the top heaviest metal pollutants in China. However, sodium-ion batteries are a potential alternative to lead-acid batteries, and they are promising in terms of environmental aspects \cite{Sodium}.

\subsection{Hardware transportation}

The production of hardware will not take place on-site, hence the emissions due to their transportation have to be taken into account. We assume that the detection units, which weigh approximately 126\,kg, will travel approximately $1700$\,km from the factory (located in Xi'An, Shaanxi Province, China) to the final site (close to Dunhuang, Gansu Province, China). The exact type of  vehicle to be used has not been decided yet; however, as there is no major infrastructure for freight transportation connecting these cities except road, the transportation will most likely  happen by truck. We chose an emission factor of 0.07\, kgCO$_2$e/tkm for truck transportation. This is an average value from the ADEME Carbon Data Base \cite{ADEMECarbonDatabase} and only serves as an order of magnitude estimate. 

For GRANDProto300, the total weight is 37.8\,t. The total transportation of GRANDProto 300 will emit around 4.5\,tCO$_{2}$e.

\section{Emission projections of the whole GRAND project}\label{section:projections}

In the previous sections we presented the emissions over the period $2015-2020$ in order to define the main sources of GHG emissions of the GRAND project. In this section, we present the estimates for the upcoming years and the next stages of the project. The estimates are computed in a {\it Business as usual} scenario, assuming that no specific actions are taken by the collaboration to reduce its carbon footprint, and that the number of collaborators in GRAND increases as presented below, and in Fig.~\ref{fig:roadmap}. This scenario allows us to compare the impact of different actions that could be implemented to mitigate the GHG emissions. 

The projections  bear large uncertainties, and should be viewed as order-of-magnitude estimates. Besides, even though we try to reflect the reality of the evolution of the experiment, unforeseen changes may occur in the future.

In the projection calculations, two types of emission sources stand out: those depending on the number of GRAND collaboration members (e.g., collaboration meetings, electronic devices), and those depending solely on the project (e.g., hardware, data transfer). 

The growth of the GRAND collaboration is foreseen to reach $\gtrsim 400-1000\,$members (see Section~\ref{section:roadmap}). As we see in the following section, we find that these numbers only have a limited influence on the total GHG emissions of the GRAND project. We thus present the upper limits derived from the 400 member-scenario in the next sections.

The three main stages of the experiment, GRANDProto300, GRAND10k, GRAND200k, as pictured in Fig.~\ref{fig:roadmap}, are assumed to last for 5, 10 and 10 years respectively. This last figure is conservative, as most large-scale ground experiments run for more than  10 years (e.g., the Pierre Auger Observatory \cite{Auger}). In our calculation of the emissions for each stage, we only take into account the emissions due to the deployed hardware and the emissions occurred during the years of operation. The yearly emissions from the collaboration members prior to 2021 are not taken into account.

\begin{figure*}[p!]
    \centering
    \includegraphics[width=0.24\linewidth]{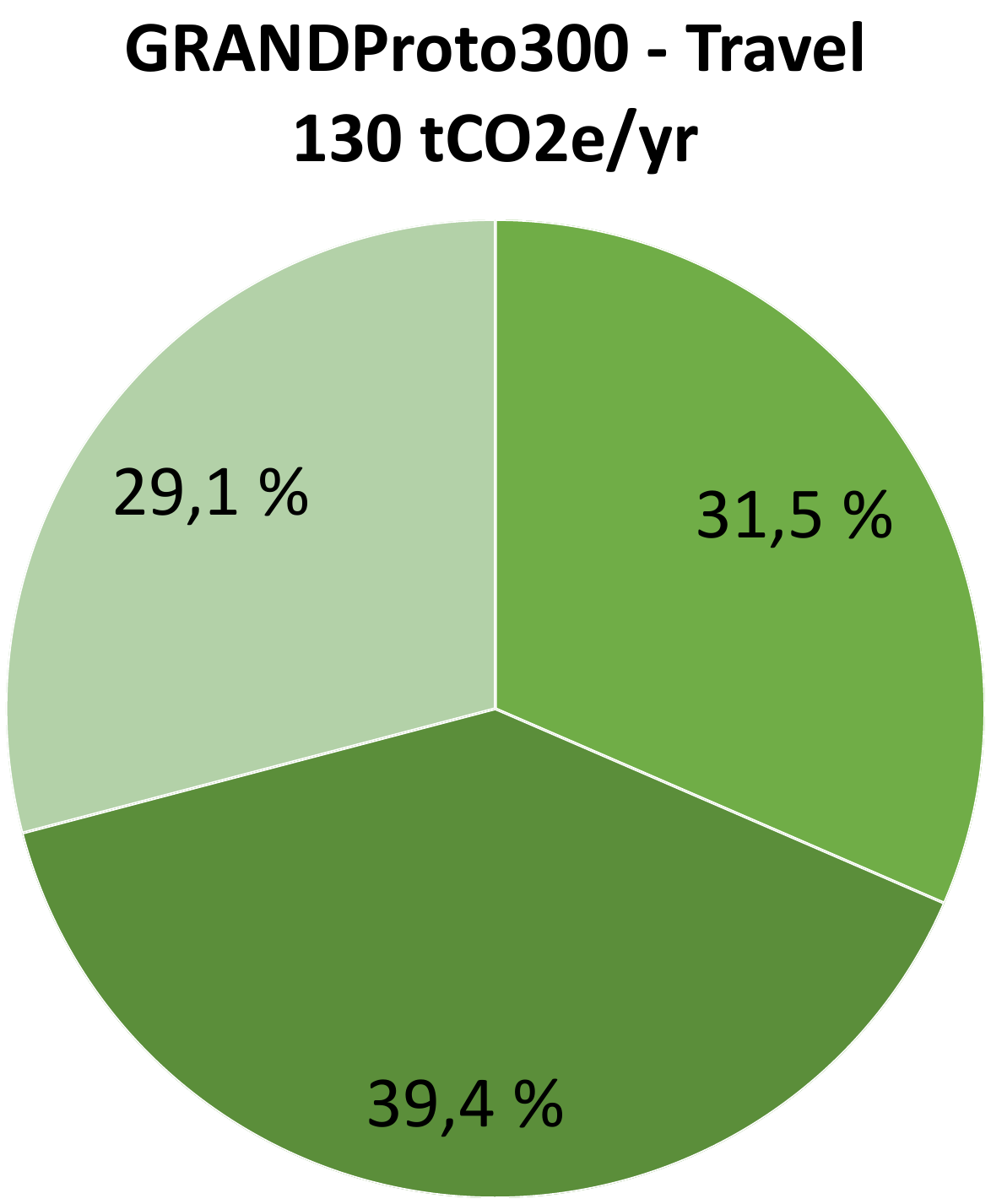}
    \includegraphics[width=0.24\linewidth]{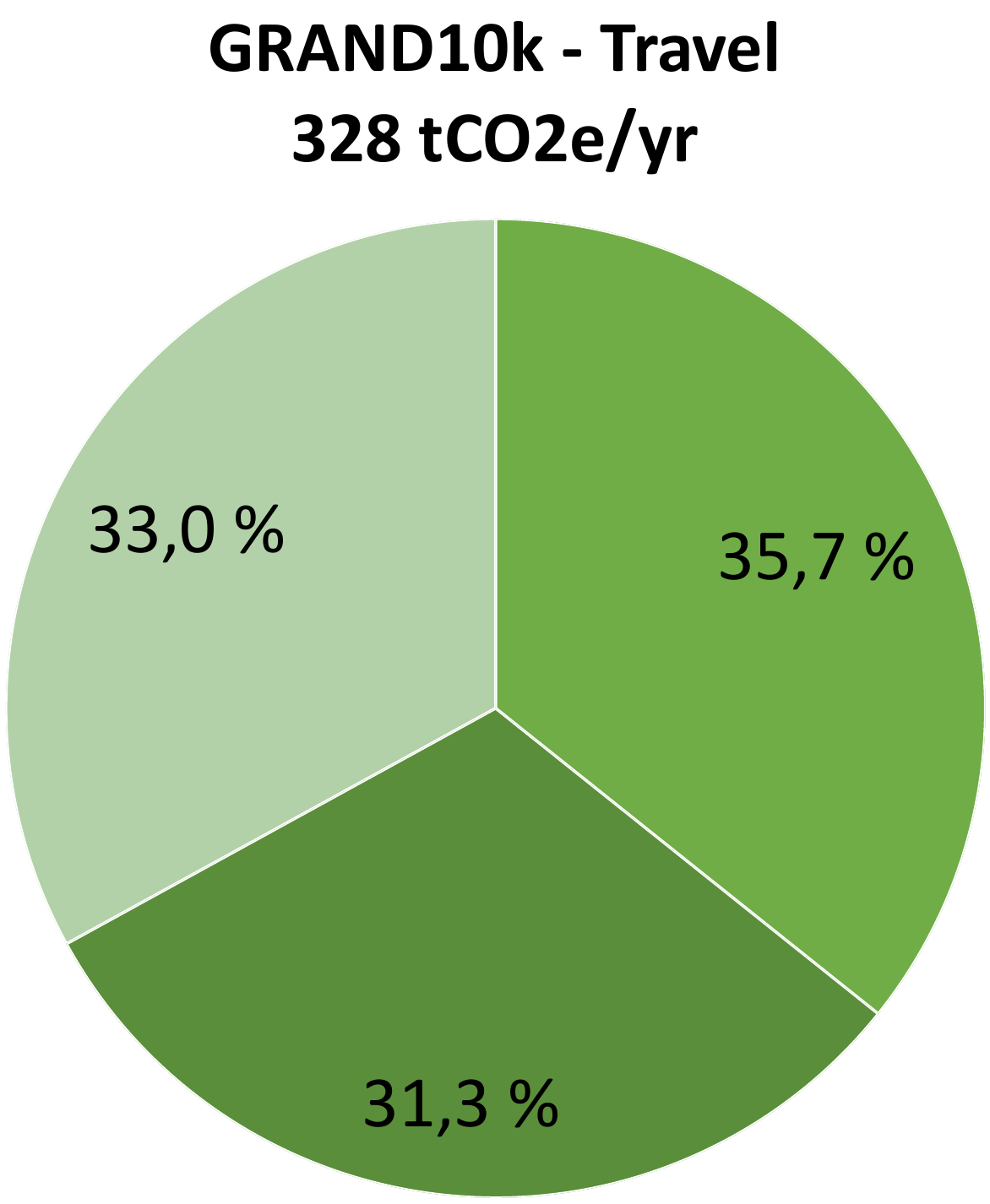}
    \includegraphics[width=0.24\linewidth]{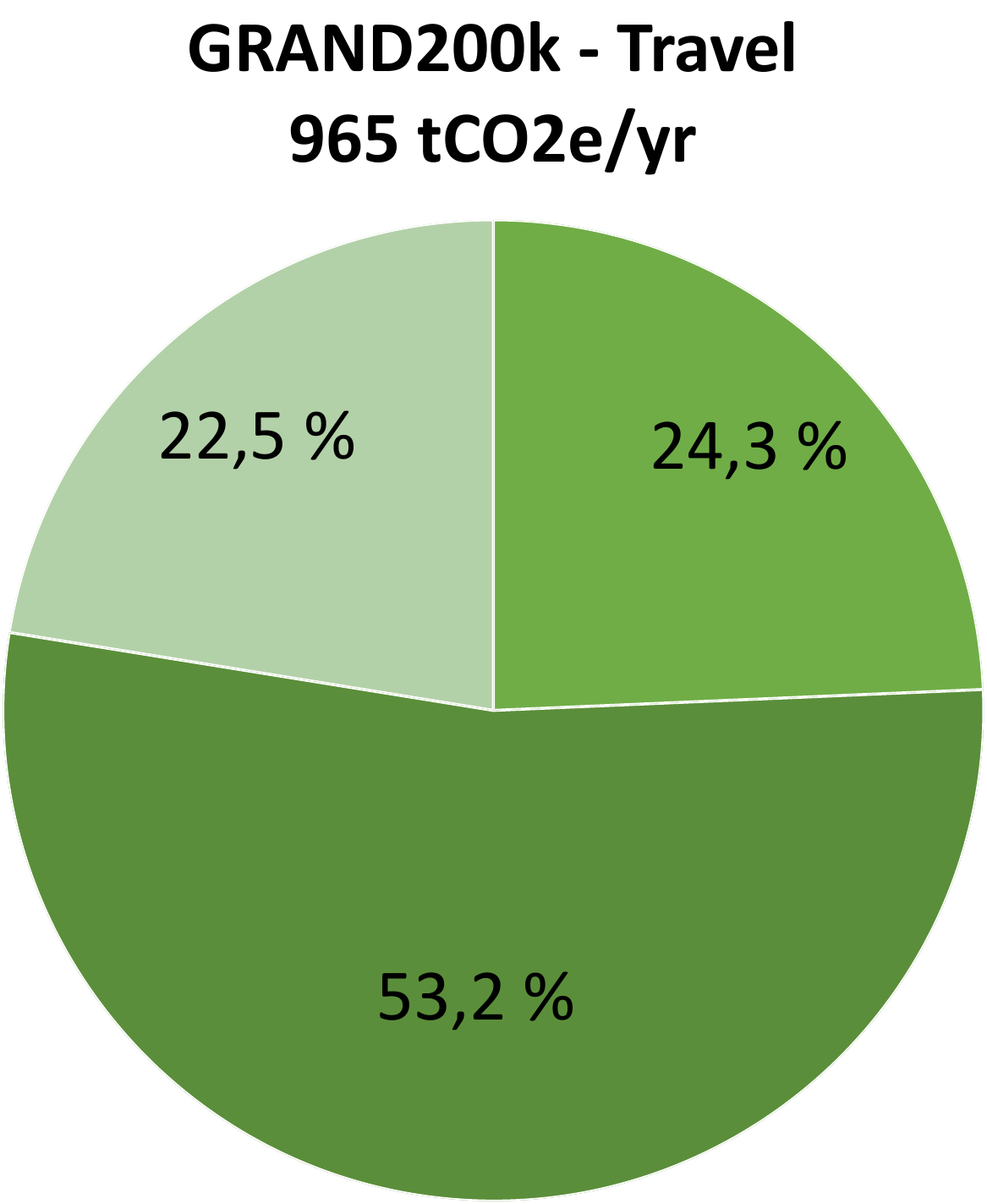}
    \includegraphics[width=0.24\linewidth]{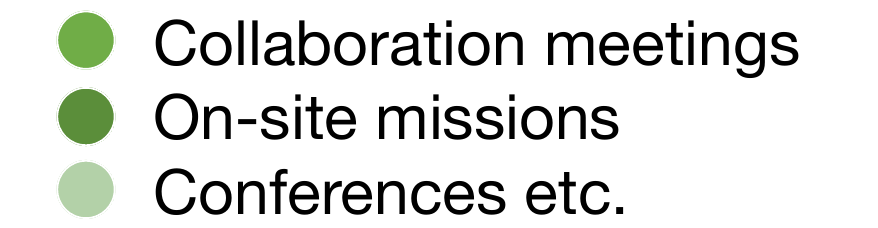}
    \vspace{0.5cm}
    
    \includegraphics[width=0.24\linewidth]{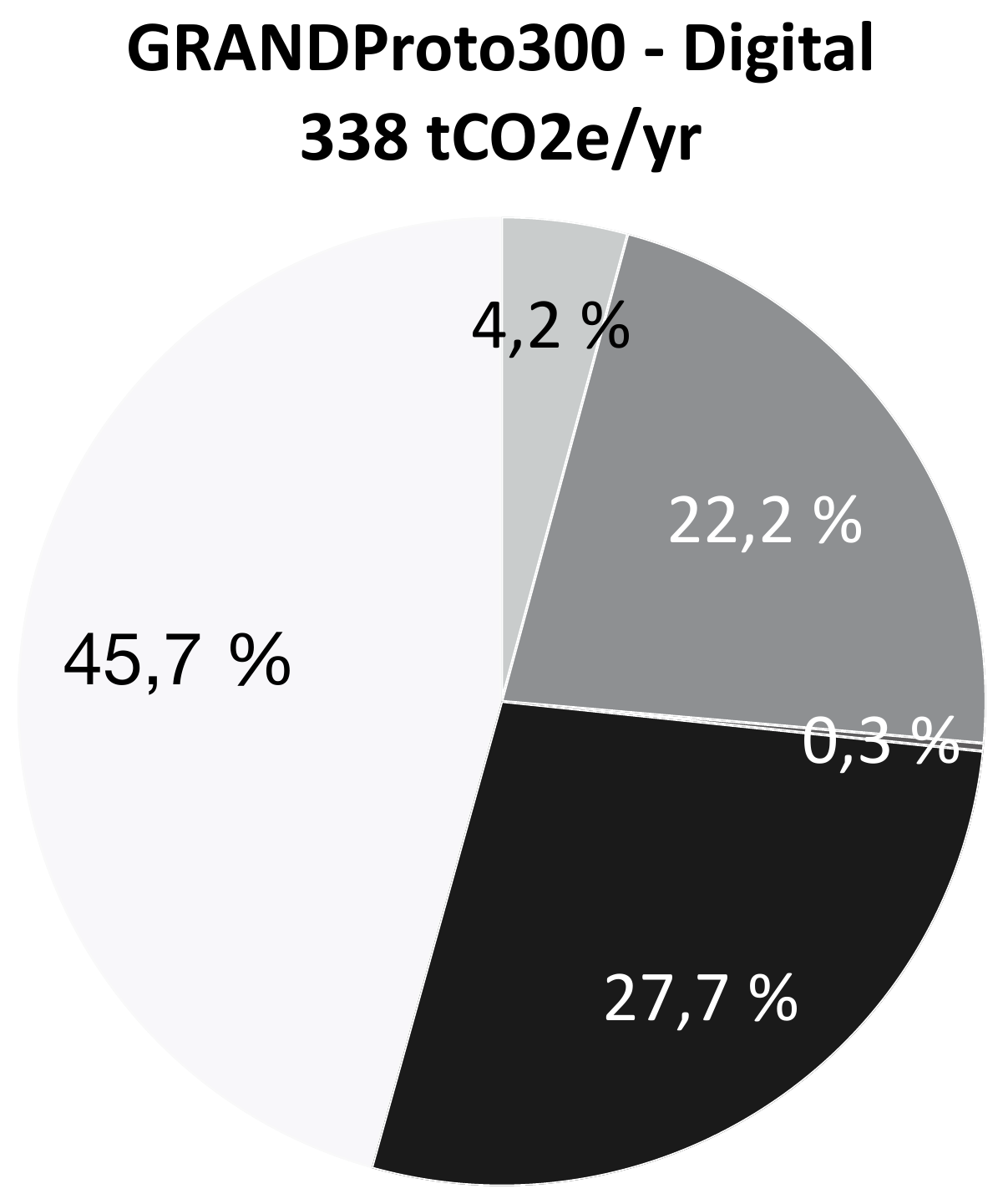}
    \includegraphics[width=0.24\linewidth]{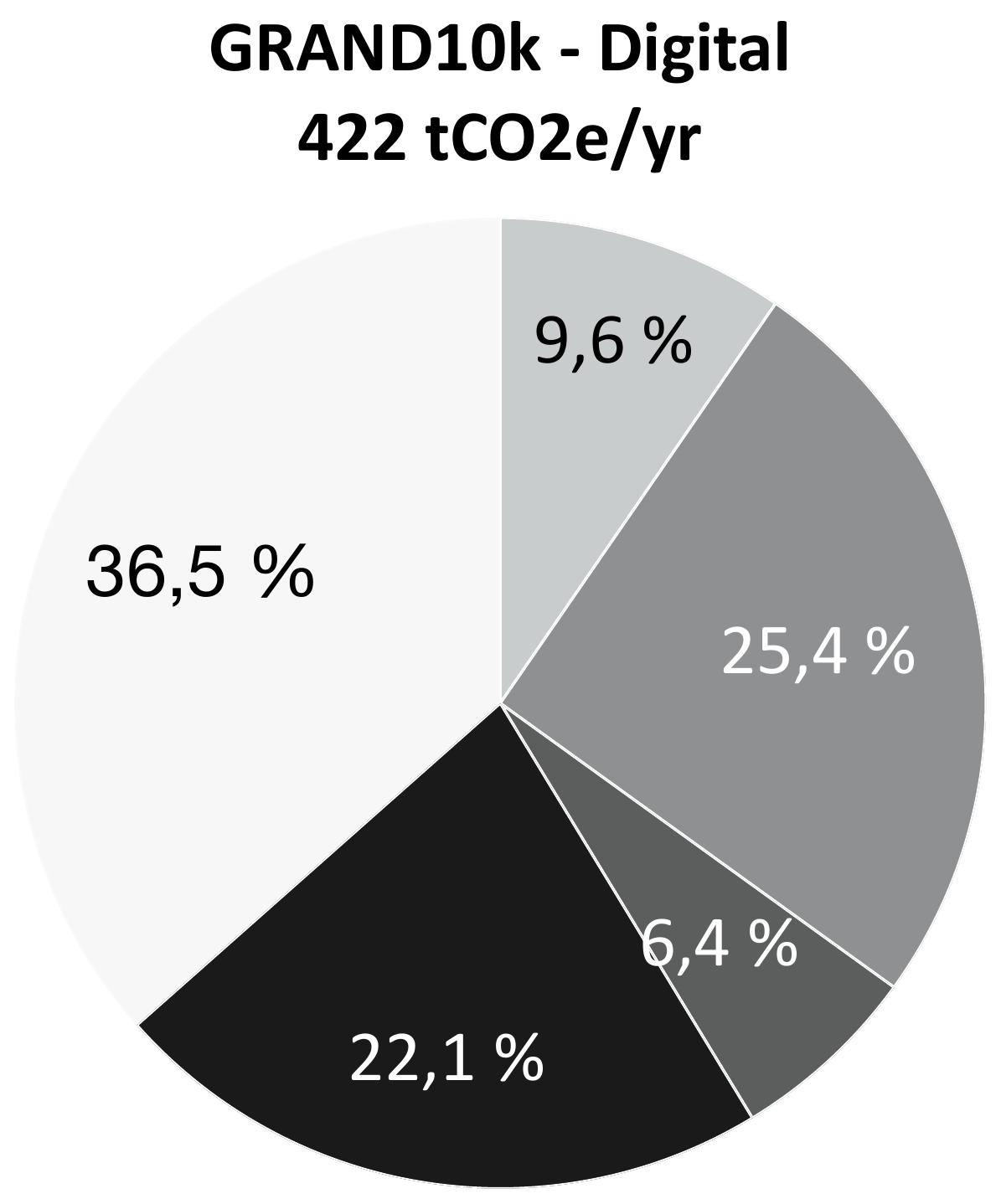}
    \includegraphics[width=0.24\linewidth]{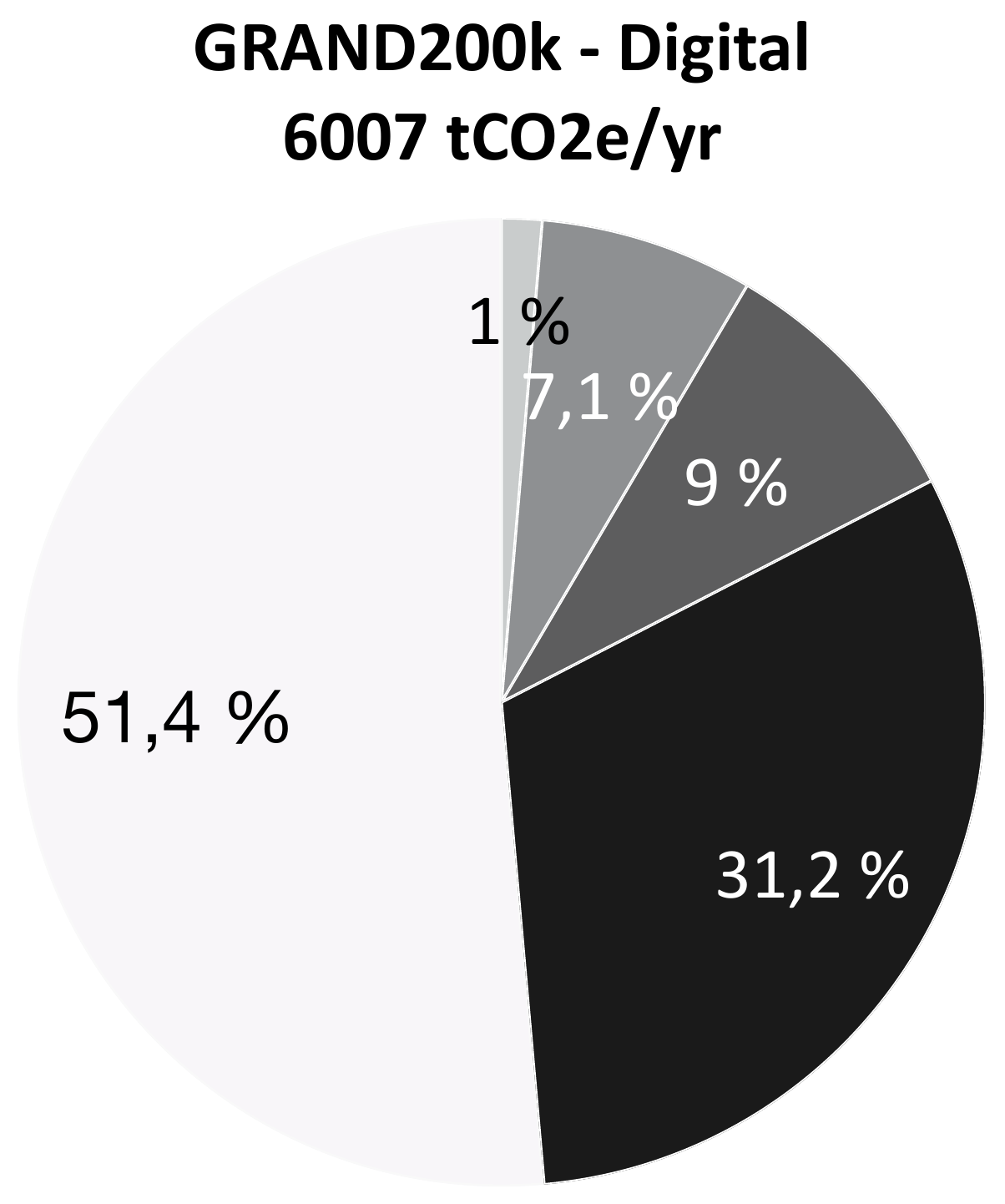}
    \includegraphics[width=0.24\linewidth]{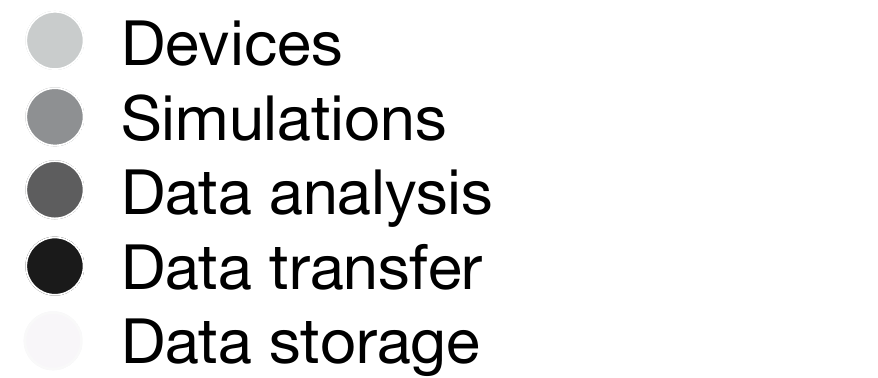}
    \vspace{0.5cm}

    \includegraphics[width=0.24\linewidth]{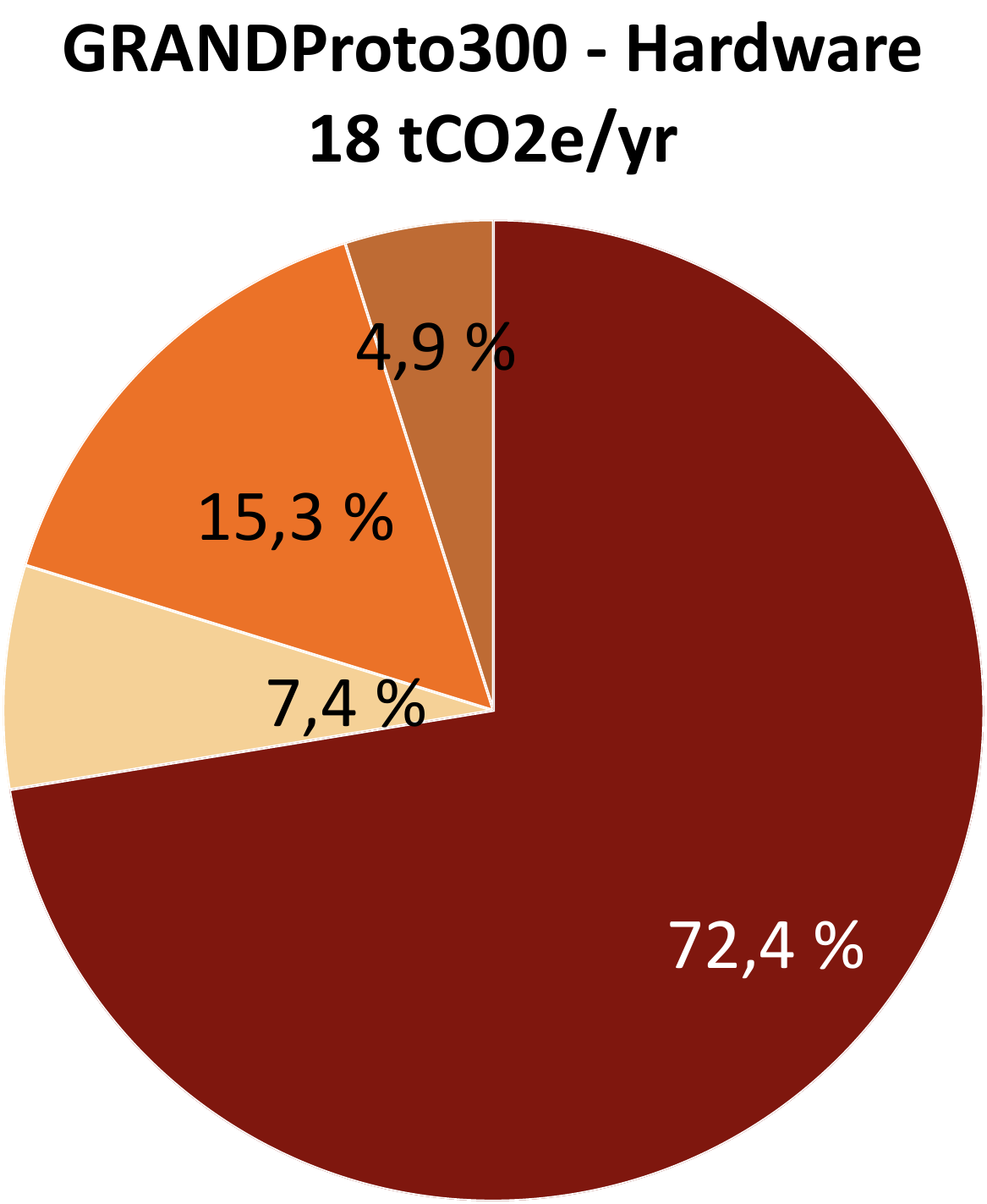}
    \includegraphics[width=0.24\linewidth]{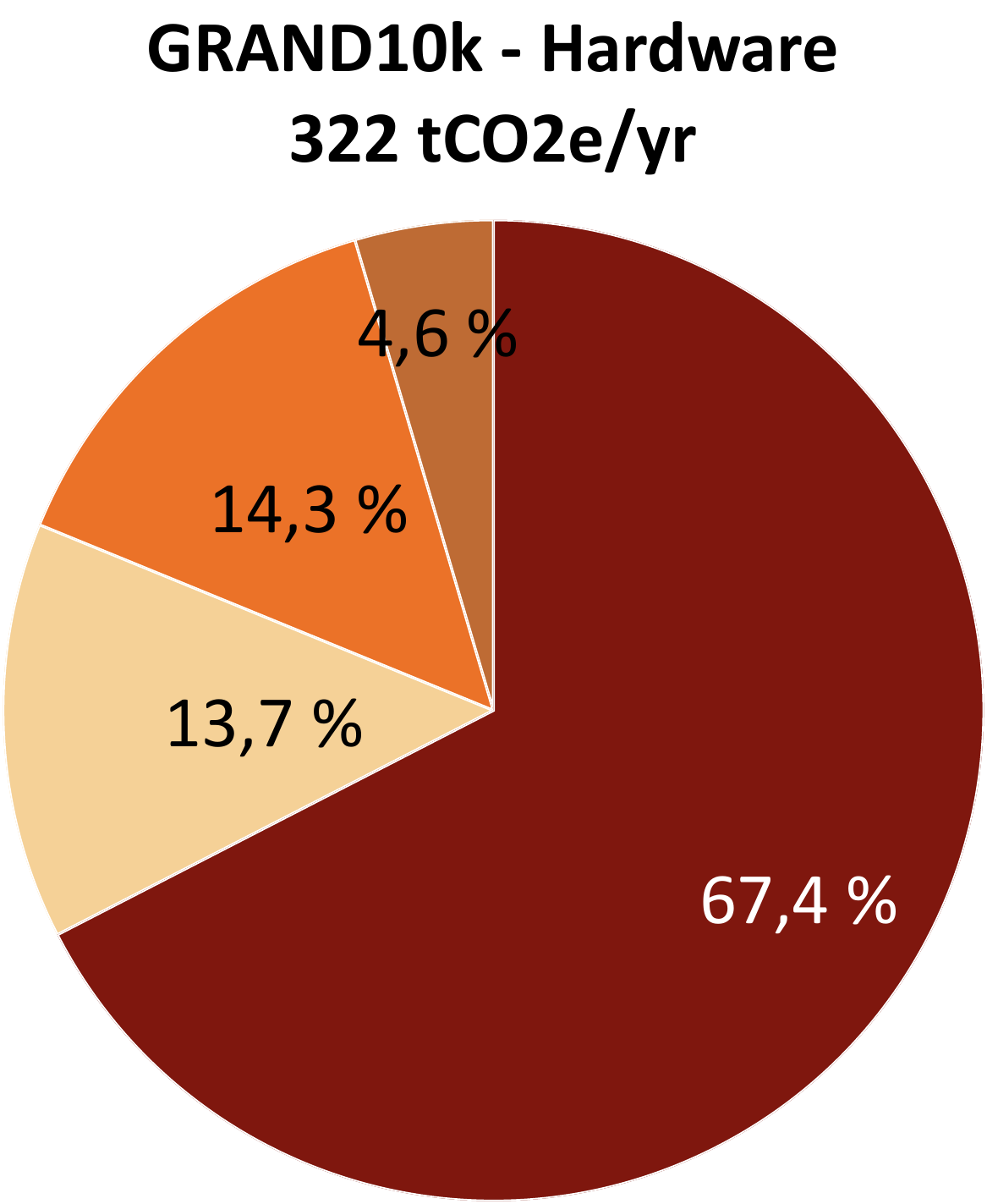}
    \includegraphics[width=0.24\linewidth]{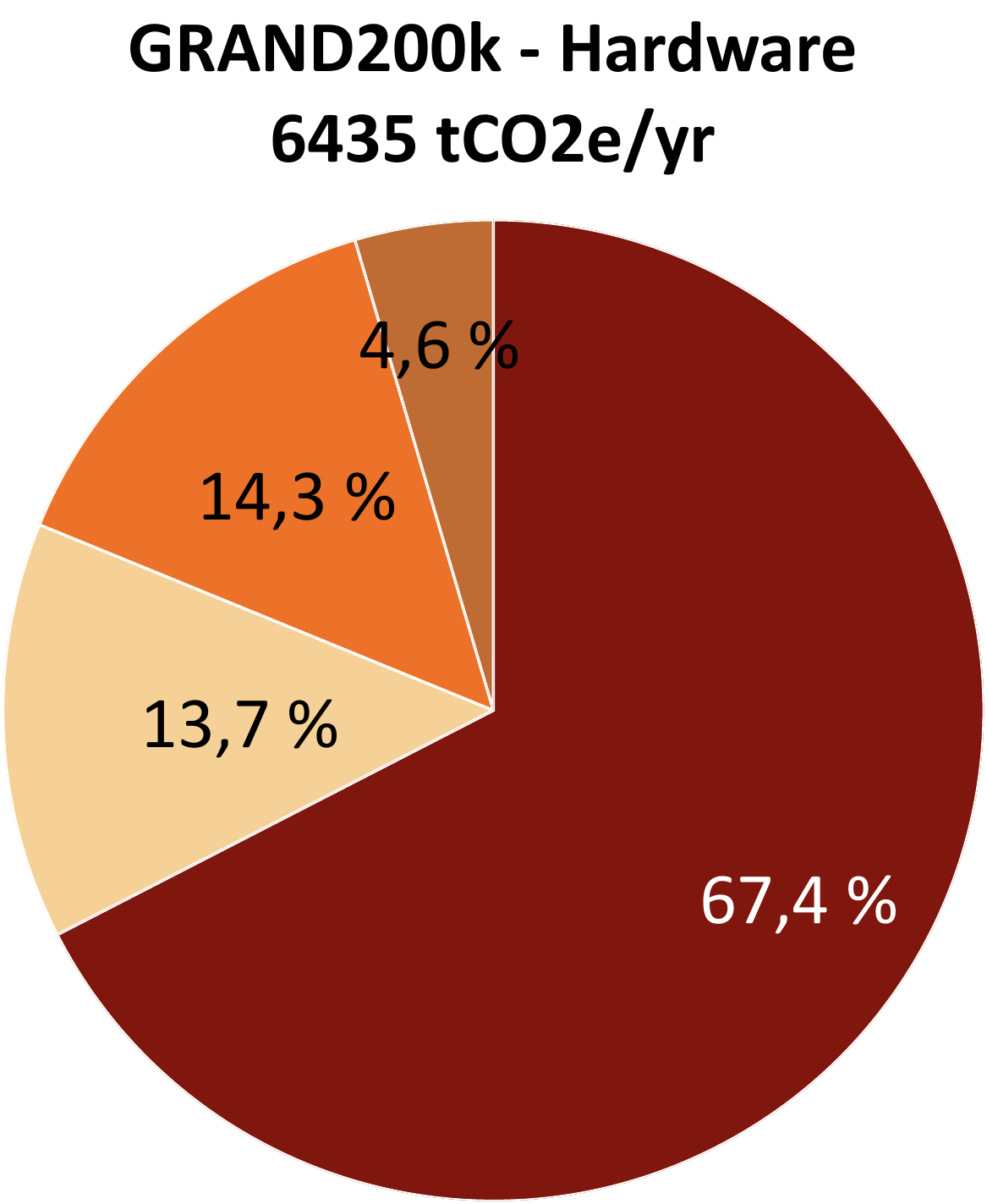}
    \includegraphics[width=0.24\linewidth]{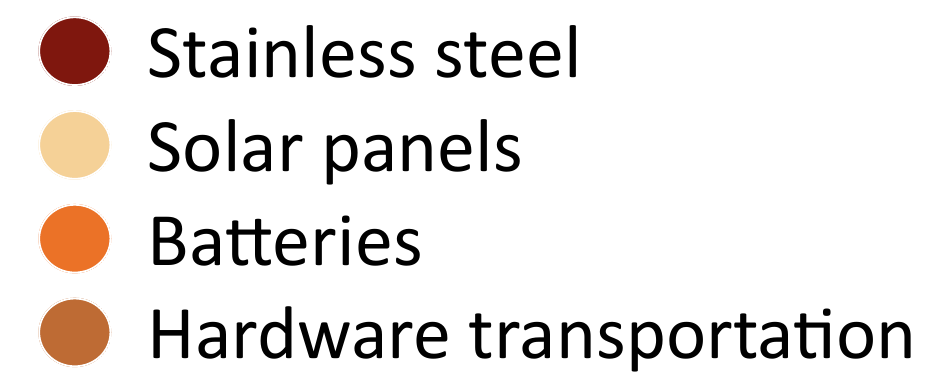}

    \caption{Projected distribution of GHG emissions related to transportation ({\it top}), digital ({\it middle}), hardware ({\it bottom}) for GRANDProto300, GRAND10k and the full GRAND array. The title of each graph indicates the total amount of emissions due to each source and each experimental stage.  These emission rates should be viewed as orders-of-magnitude estimates due to the large uncertainties inherent to projections and to emission factor estimates.}
    \label{fig:proj_3sources}
\end{figure*}

\subsection{Travel} \label{section:proj_transportation}
The emissions related to collaboration meetings and "other" purposes (e.g. participation to conferences, visits) identified in Section~\ref{section:transportation}) will grow roughly proportionally to the size of the collaboration. We estimate the yearly emissions per person in the collaboration for collaboration meetings and ``other" purposes, by calculating the average emissions per person and per year over the period $2015-2019$. Emissions due to travelling to collaboration meetings and other purposes amount respectively to  0.6\,tCO$_2$e/person/yr and 0.5\,tCO$_2$e/person/yr . Those numbers are calculated on average and take into account every collaborator of GRAND. The emissions for the three main stages of the experiment can then be calculated by multiplying these rates by the corresponding number of members and the number of years. 
 
The emissions related to on-site missions do not depend on the number of collaborators, as only a limited number of members are needed for on-site missions, whatever the stage of the project (GRANDProto300, GRAND10k or GRAND200k). For on-site missions, the emissions are more influenced by the different experimental stages: in particular by the number of units deployed , the number of sites, and their distance from one to the other. On-site missions are also expected to have a higher impact on the carbon footprint at later stages  as the different sites for GRAND200k might be located on different continents. Due to the current high uncertainty related to the future location of the site, we chose not to take that effect into account.

We estimate  yearly emissions of on-site missions by averaging the emissions calculated over the years $2015-2019$. The yearly rate corresponds to 51.3\,tCO$_2$e/year for the collaboration.  In order to derive projections for the three stages of GRAND, we assume that this benchmark rate will remain constant during the years of operation of GRANDProto300, will double during GRAND10k, and will be multiplied by 10 during GRAND200k. The GRAND10k phase will roughly be an extension of the existing GRANDProto300 site, hence no drastic increase in on-site missions is expected . At the final stage of the project, about half of the $10-20$ sub-arrays will be deployed in China and the rest will be deployed worldwide. Adequate experimental sites will have to be searched, implying a large increase in on-site missions.  

These assumptions lead to the proportions presented in the first row of Fig.~\ref{fig:proj_3sources}. One can notice that each of the three travel purposes take up about 1/3 of the travel emissions for the first two phases of the project, whereas on-site missions prevail in the last phase.

These pie charts show the importance of taking actions on all three items in order to reduce the impact of travelling. The logistics should also be carefully planned in the global GRAND200k stage, in order to minimize the number of on-site missions.

\subsection{Digital technologies}\label{section:proj_digital}

The emission projections presented in this section are highly dependent on technological change. We present the proportions of emissions related to the different digital sources for the three main stages of the GRAND project, in the middle row of Fig.~\ref{fig:proj_3sources}. Below, we mention the assumptions made in order to calculate these projections.

\subsubsection{Electronic devices}
Projections for electronic devices are proportional to the size of the collaboration. For projections, we use the emission factors presented in Part \ref{section:digital}. In the {\it Business as usual} scenario, we suppose that the lifetime of the devices is still the shortest presented in Table \ref{tab:devices}. The middle row of Figure~\ref{fig:proj_3sources} shows that the contribution of electronic devices remains limited among the digital emission sources. This stands even when assuming a strong increase in the collaboration size.

\subsubsection{Simulations and data analysis}
Projections for numerical simulations and the related use of CPU time are difficult to make as they are  dependent on the way scientists carry out research. Making projections for future years is highly hypothetical and uncertain. Moreover, the electricity intensity of Internet data transmission is dependent on the year, as digital technologies are changing fast. Therefore emissions computed in 2020 will have to be revised later in order to have a more accurate estimate, as the project will be fully deployed in the 2030s. 

There are several factors that can influence the future use of CPU time for simulations: the number of local groups that run simulations for their own research, the first results from GRANDProto300 that will call for more simulations (or not), the level of accuracy needed for the simulations, the number of people that will work on those simulations, etc. 

The current members involved in the GRAND simulations estimate that about 6\,M\,hours CPU will be needed overall for GRANDProto300. For the next phases, it is likely that the same library can be used, resorting for instance to interpolation and semi-analytical tools that are being developed \cite{Zilles20_RM, tueros2020synthesis}. How much CPU will be used (as well as how many new simulations will be needed) would then depend more on how many studies are being pursued (i.e., how many people work on the project) rather than on the number of antennas in the array. We will hence start from the hypothesis that the GHG emissions from simulations correspond to 26.9\,tCO$_{2}$e (6\,M\,hours CPU) for GRANDProto300 and scale with the size of the collaboration for the next stages. More discussions can be found in Section~\ref{section:proj_simulations}. \\

For data analysis on the other hand, one expects that the volume of data to be analyzed would scale along with the size of the experiment (i.e., the number of antennas). Indeed, the signal extraction requires the full data to be scanned and the corresponding CPU-time will scale with the data volume. The further layers of more "science-driven" data analysis would scale with the number of collaboration members, but this can be considered as a second-order effect. In order to make projections, we have scaled the GHG emissions from data analysis in GRANDProto300 (see Section~\ref{section:simulations}) to the GRAND10k and GRAND200k phases.

\subsubsection{Data transfer and storage}\label{section:proj_data}

As shown in Section~\ref{section:digital} data transfer and storage is an important source of GHG emission for the GRAND project, especially if raw data is transferred and stored. The data volume in the future will mainly scale with the number of detection units that will be deployed, which are already defined for each step of the project (300, 10k, 200k for GRANDProto300, GRAND10k, GRAND200k respectively). This yields a gigantic volume of 2\,EB/yr (exabytes/yr) of raw data to store and transfer for the 200k antenna stage. 

Starting from the 10k phase, a data center will be built on-site in order to directly process and reduce the data storage. For instance, for the radio-astronomy data, by adding the signals incoherently over the antennas (as opposed to performing heavy Fast-Fourier-Transform-type operations to phase the signals) and storing only the sum (as opposed to keeping the individual signal for each antenna), we can divide the volume of data by the number of antennas, and expect to gain 4 orders of magnitude in volume for GRAND10k.  For GRAND200k, the gain would also be of 4 orders of magnitude, as it is limited by the size of each sub-arrays (of $\sim 10^4$ antennas each), over which the information can be added.

As done in other radio experiments with large amounts of data processing, the data will be classified according to the treatment they have undergone. Raw data (D0) will be quickly processed and erased after a few months. A  class of data with a first layer of reduction (D1) will be stored in a cloud until the end of a given observation program, and another class will be archived (D2). The reduction from D0 to D1 will thus happen on-site, D1 to D2 will be done in a nearby data center, in the country of each 10k sub-array, and only D2 will be transferred worldwide. 

In practice, D2 will mostly correspond to the data "necessary for astroparticles" that were mentioned at the end of Section~\ref{section:data_transfer}. The data to be transferred and stored will then be considerably reduced to $1\,$PB/yr for GRAND10k and $20\,$PB/yr for GRAND200k compared to a simple scaling from GRANDProto300 emissions.  

We calculated our estimates for the GRAND10k and GRAND200k stages of the project assuming that the reduced data will be transferred and stored in 3 different places. 

Figure~\ref{fig:proj_3sources} shows that even with data reduction at GRAND10k and GRAND200k stages, data transfer and storage are the major sources of GHG emissions among the digital emissions.

A great caution should be applied however to the emissions projections presented, as the electricity intensity of data transmission changes upon time along with technology changes. For example, Ref.~\cite{InternetAslan} finds that the electricity intensity of data transmission has decreased by half approximately every 2 years since 2000. One should also keep in mind the emission goals set by the Paris  agreement will have to cause a significant change in the GHG emission of our electricity almost anywhere in the world. For conservative purposes, we have not taken those changes into account. 

From another perspective, the construction of an on-site data center will imply large emissions to bring that much electric power in such a remote place. The emissions related to the center have not been estimated as the type and size of the data center are currently unknown. 

\subsubsection{Hardware equipment}\label{section:proj_hardware}

The projections for hardware equipment are proportional to the quantity of units deployed. Each unit will likely be lighter than the current ones, which will lead to the use of less raw material such as stainless steel (see Section~\ref{section:action_plans}). Still, this potential trend is not taken into account in our projections, as we always choose the more conservative option.  Figure~\ref{fig:proj_3sources} shows that stainless steel is by far the most emitting item among hardware equipment and their transportation. The emission proportions are almost identical throughout the 3 stages of the project, which results from the scaling that we have performed on the detection unit numbers. 

\subsubsection{Hardware transportation}

For the final stage of the project, where sub-arrays will be located throughout the world, the transportation of the antennas over different continents will also have to be taken into account, unless most of the production can be done in the country hosting the sub-array.

Realistic projections are difficult to make as the location of the sub-arrays will only be decided in the coming years. To give an indication, we have extrapolated the hardware transportation emissions of the GP300 phase to 200k antennas, assuming a local production and that the units would all be hauled by truck over distances similar to Xi'An-Dunhuang. 
For GRAND10k the emissions will amount to 149\,tCO$_{2}$e and for GRAND200k it will represent 3,000\,tCO$_{2}$e.

 The extrapolation is proportional to the number of units, as transportation emissions  are proportional to the quantity of units hauled. Even though the different sites of GRAND200k will be around the world, we assume that the distance traveled by the equipment will be the same on each continent and be covered by truck. As a consequence this extrapolation incorporates high uncertainties in terms of distances and modes of transportation.

\subsection{Overall emissions and their distribution}

\begin{figure*}[tb]
    \centering
    \includegraphics[width=0.24\linewidth]{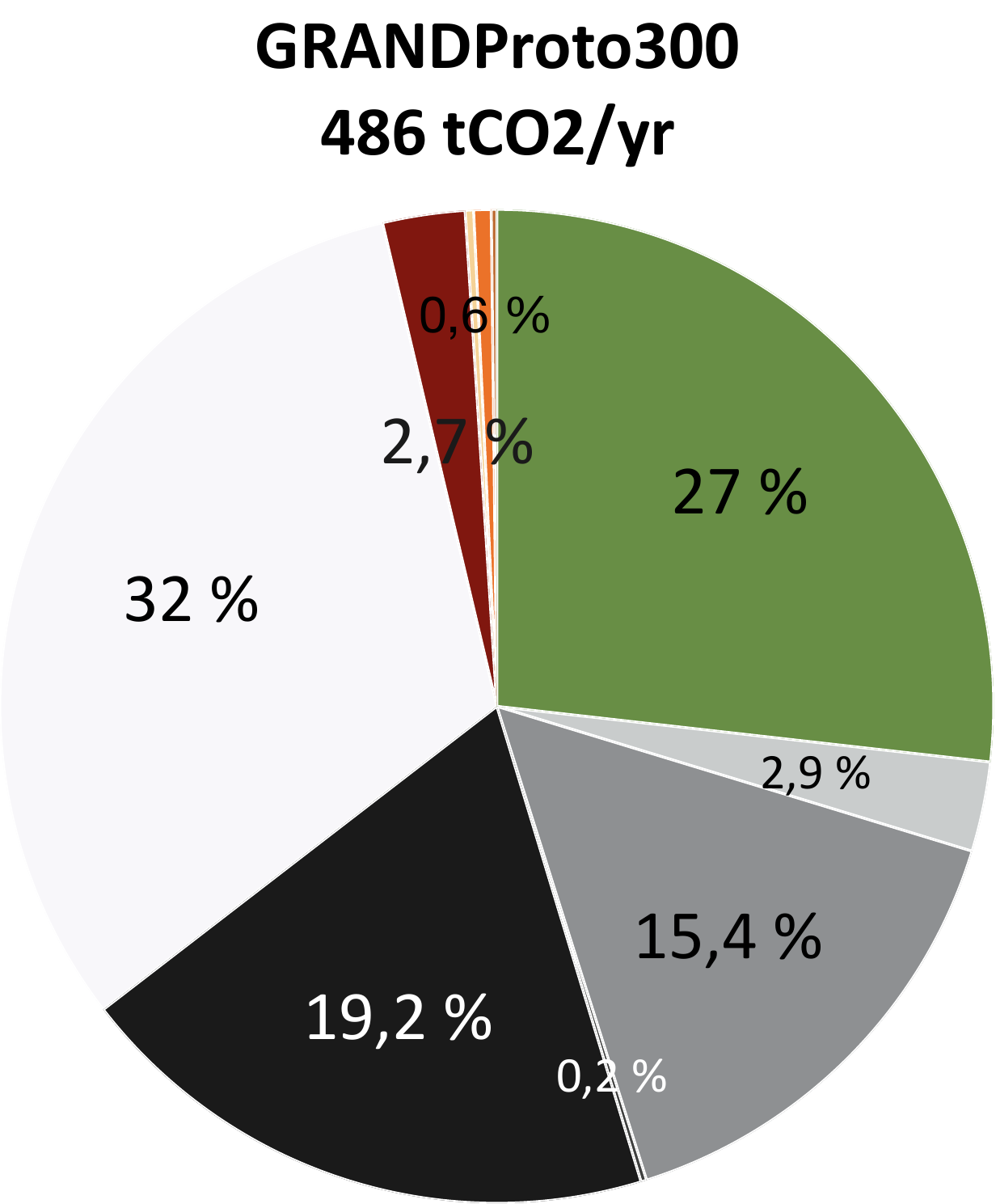}
    \includegraphics[width=0.24\linewidth]{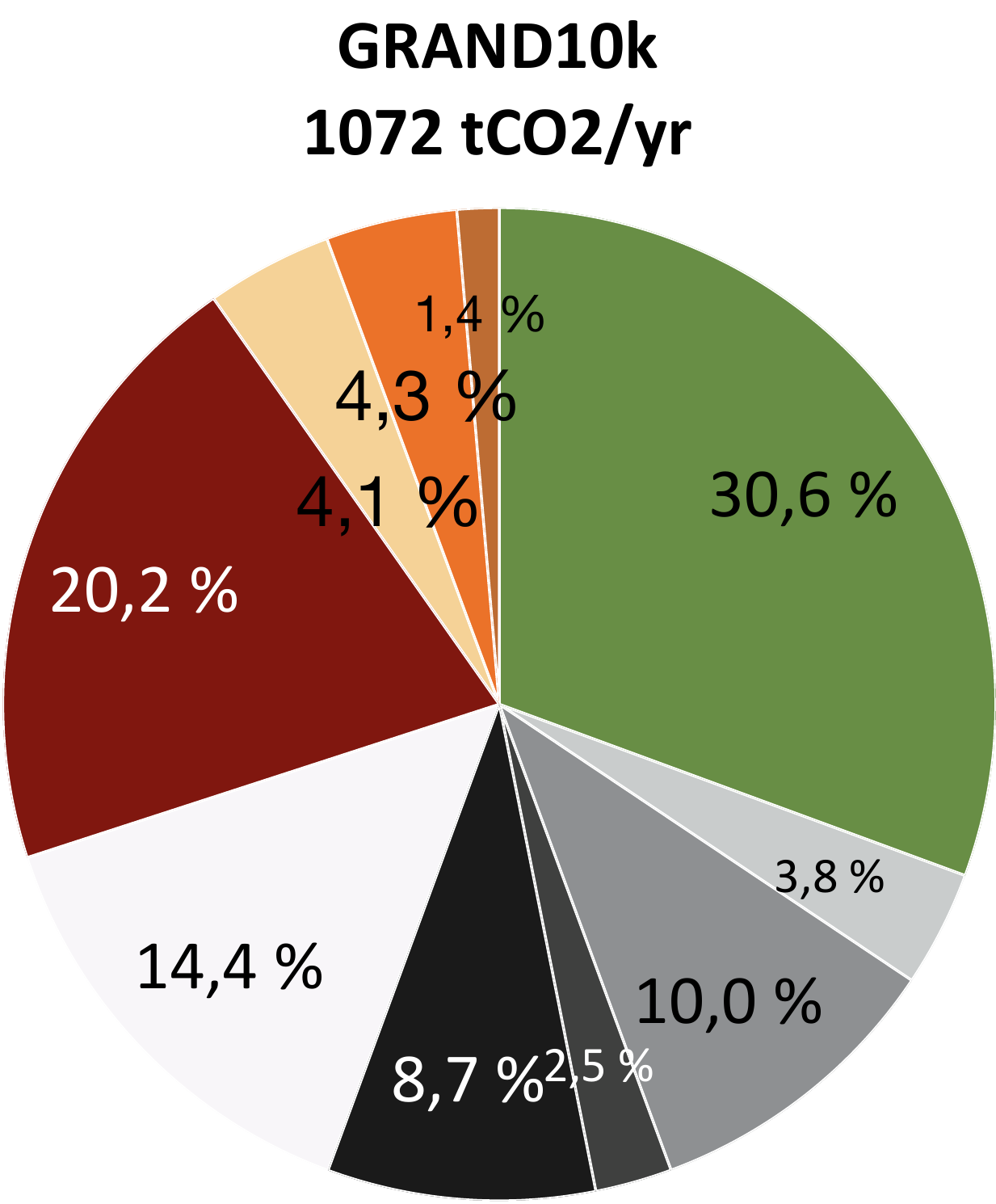}
    \includegraphics[width=0.24\linewidth]{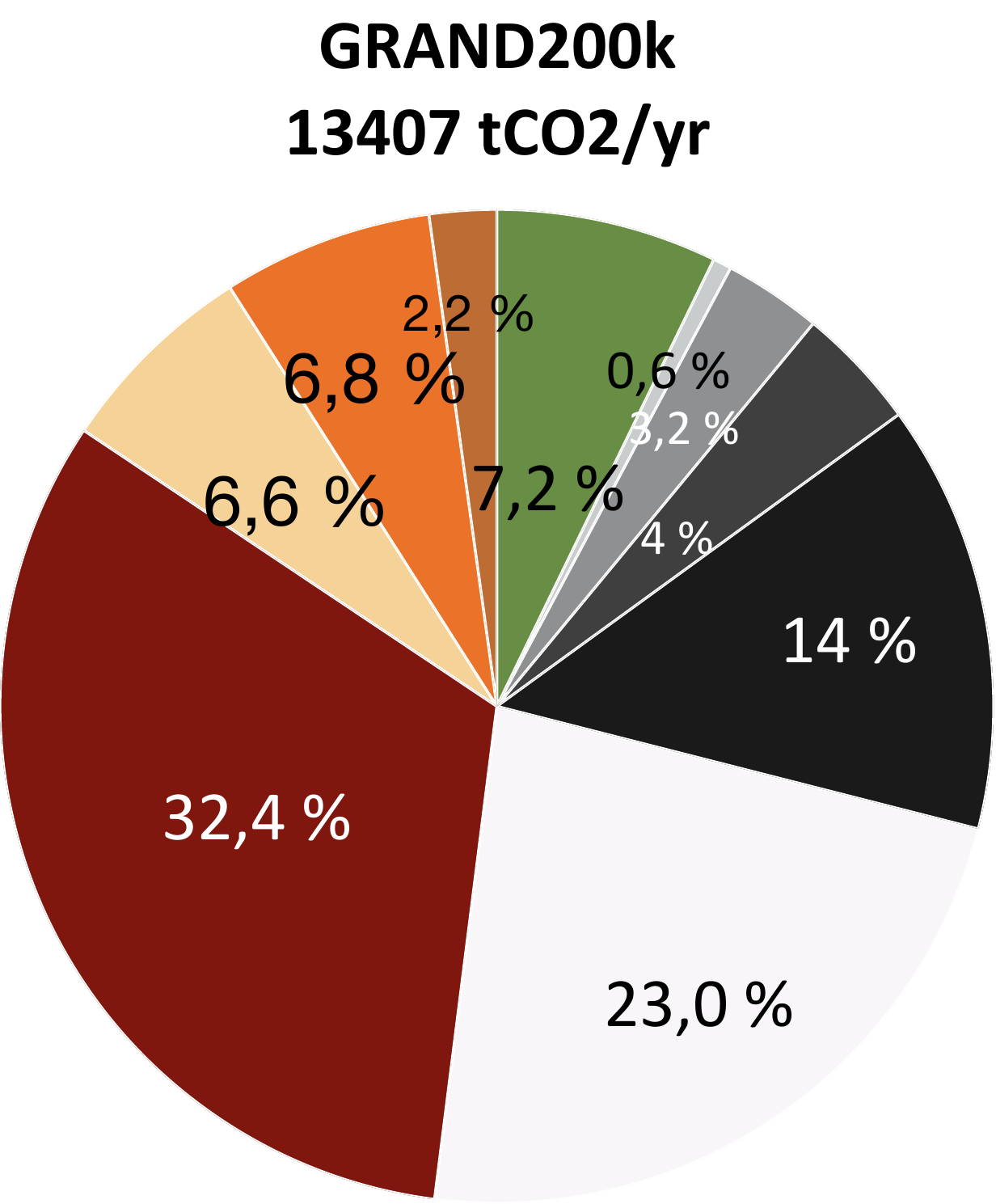}
    \includegraphics[width=0.24\linewidth]{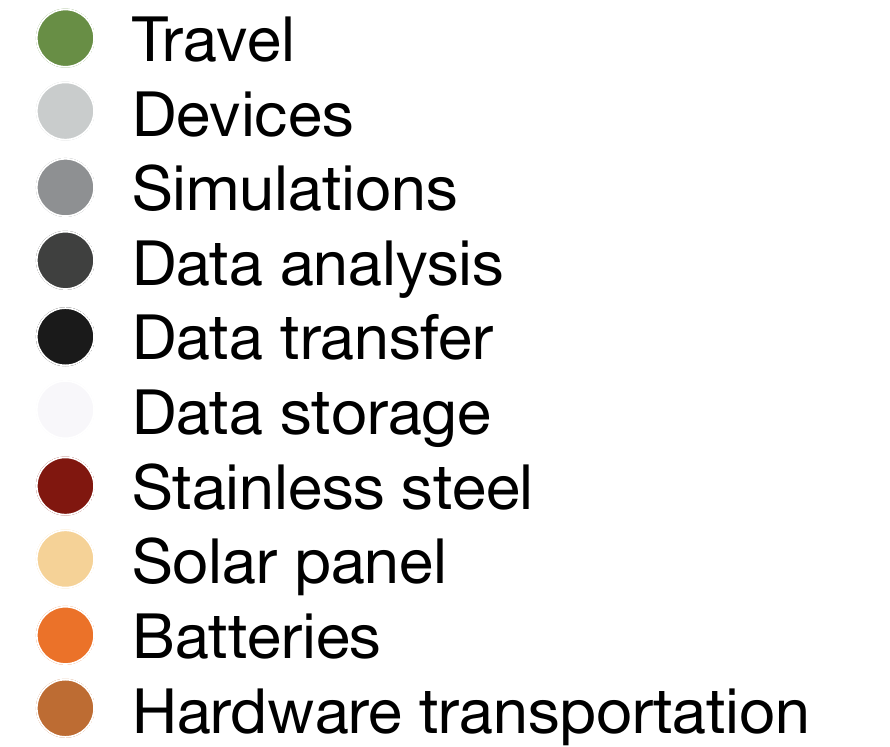}
    \caption{Projected distribution of GHG emissions for all sources for GRANDProto300, GRAND10k and the full GRAND array. The title indicates the total amount of emissions per year due to each source at each experimental stage. These emission rates should be viewed as orders-of-magnitude estimates due to the large uncertainties inherent to projections and to emission factor estimates.}
    \label{fig:proj_all}
\end{figure*}

We merge the projections obtained for the three types of emissions (travel, digital, hardware) and present a global distribution of the emissions in Figure~\ref{fig:proj_all} for the three stages of the GRAND project. 

These three stages have highly different emission scales, with two orders of magnitude more emissions at the final GRAND200k stage, compared to the prototyping stage. 
This results from the sheer size of the experiment (the number of detection units). As an illustration, the 13\,400 tCO$_2$e/year emission estimate of the GRAND200k phase represent about 7900 Paris-Dunhuang return flights. Another comparison can be made with car manufacturing, which emits roughly 15\,tCO$_2$e per car \cite{Car_manufacturing_EC}: the emissions from GRAND200k per year corresponds to that of the production of less than 1000 cars.

One can notice the shift in the prevalence of the various emission sources (travel, digital and hardware) for the three stages of the project. For the small-scale prototype GRANDProto300, digital emissions prevail followed by travel, while for GRAND10k, the three types of sources become equally important. Finally, for the final GRAND200k stage, hardware equipment prevails, stainless steel being the major emitter.

\section{Action plans}\label{section:action_plans}

In view of the results presented above, several actions were discussed within the GRAND collaboration in order to limit the GHG emissions of the project. In the future, a strong focus will be put on the major emission sources, namely on the use of stainless steel and on data transfer and storage. However, as of today, a focus should also  be put on travel, as it constitutes one of the main emission sources of the small- and mid-scale stages of the project. Besides, it is our belief that mitigation measures should be taken on all possible fronts. Therefore action plans are discussed on most items evaluated in this study, as detailed in the next sections. Following these ideas, the collaboration will aim at establishing in the near future a GRAND Green Policy, which each member will be encouraged to follow to reduce the collective carbon footprint.

\subsection{Travel}

\subsubsection{On-site missions}

On-site missions constitute the prevailing emission source among other travels related to GRAND. Although it might be the item on which one has the least leverage, as they are essential to the experimental operations, there is some room for actions. 

 While the experiment is being designed, deployed and tested, a handful of international specialists are required to go in the field and work on the equipment. However, once the experiment is operational, routine checks and the resolution of technical problems can be performed by a dedicated technical staff.

Hence, one way to reduce the distances traveled would be that most of these on-site missions be performed mainly by colleagues from Chinese institutes, who work closer to the site. On-site missions could also last longer, with possibly engineers living several months in the nearest city (Dunhuang, Gansu Province). If they cannot stay longer due to experimental/weather/personal reasons, these travels could be combined with other missions such as visits to Chinese institutes, small group workshops, collaboration meetings, conferences. 

Our {\it Business as usual} scenario hypothesis is that the number of on-site missions would be scaled up by an order of magnitude for GRAND200k. It is highly unclear whether this number is realistic, as it will depend on the involvement of the members of each country where the sub-arrays will be installed. If the on-site missions can be mostly performed by local colleagues, this could drastically reduce the emissions of the final stage of the project.

\subsubsection{Collaboration meetings and other travels}

Setting an action plan for the other travel purposes (collaboration meetings, conferences, visits...)  is also necessary.  For students and postdoctoral scholars, networking may often be perceived as a sine qua non for a successful career. Reference~\cite{Wynes19} shows however that the amount of air traveling does not necessarily translate into academic performances.  Indeed, more and more research members agree with the necessity to reduce flying. For example, the poll carried out among a representative sample of the research community in France in 2020 reveals that almost two thirds of the respondents are ready to reduce their flights to attend conferences, meetings or congresses \cite{Resultats1point5}. The COVID-19 pandemic has led the research community to experiment virtual meetings and conferences and analyse their impacts in various respects including attendance and carbon emissions  \cite{stevens2019imperative,2020NatAs...4..823B,2020Natur.583..356K}.

In order to reduce the emissions related to conferences where GRAND results are presented, the collaboration could prioritize sending  members geographically close to the location of the conferences. One could also consider limiting the total number of conferences to be attended per year to present GRAND results. 

As for reducing the carbon impact of collaboration meetings, several ideas could be implemented: 
\begin{enumerate}[noitemsep]
\item Choosing a location which is less carbon emitting by running an algorithm knowing the geographical location of each member \cite{ResponsibleAcademia}.
\item Holding a virtual meeting one year out of two, using virtual tools such as Mozilla Hub \cite{MozillaHub} for coffee breaks.
\item Instead of a completely virtual meeting, hold a hybrid multi-hub meeting, setting up continental hubs (e.g., Europe, China, US, Brazil) where members gather, and communicate virtually between the hubs as it has already been done \cite{Organize}\cite{Low-carbon}.
\end{enumerate}
The last  two actions should help to cut the emissions due to collaboration meetings by a factor $1.5-2$.

Moreover, on a different note, collaboration meetings could be made low-waste (in particular in terms of glasses, cutlery, plates) and could offer only vegetarian buffets.

\subsection{Digital}

\subsubsection{Data transfer and storage}

Digital is one of the major sources of emissions in the project, with data transfer and storage being the most emitting ones. With the proposed scheme of data storage and transfer (Section~\ref{section:proj_data}), the collaboration is planning to drastically reduce the volume of data to be archived. 

Data transfer worldwide to the final cloud storage data centers could be reduced by hard-copying them on hard-drives and by mailing/shipping them to the centers. It is interesting to notice that flying the data by plane 4 times a year would be many orders of magnitude less carbon emitting than transferring the data online. However, while this comparison is true today, electricity in 2030+ could be produced by more renewable resources in Europe and China at least, which will have an impact on the emission factor of electricity. While the emission factor of electricity is likely to decrease, the emission factor of plane will probably remain high, and transferring data by plane could not be the best option in the next years.

In the future, it is likely that the emission factor for the electricity intensity of data transmission decreases, because electricity will be less generated by fossil fuel power plants. The cloud storage systems may become more efficient and less consuming. 

The collaboration will also work on the best solutions for data reduction, in order to store the maximum of relevant information at different levels, while reducing the volume of data to be archived. 

One idea to reduce the emissions related to data storage is to store in data centers of countries where the emission factor of energy is low. For example, in France, the emissions related to the storage of all GRANDProto300 data would be 16.8 tCO$_{2}$e, to be compared with the value of 153.6 tCO$_{2}$e presented in section~\ref{section:data_storage}, which uses the global energy emission factor estimated in \ref{section:data_transfer}.

\subsubsection{Numerical simulations}\label{section:proj_simulations}

Numerical simulations represent a large share of the emissions in the GRANDProto300 phase. In the next phases, this source becomes less important because it scales as the size of the collaboration. In order to limit the number of simulations run by different members and the related CPU time, the collaboration is setting up a common simulation data library.

On top of the possible actions already mentioned in Section~\ref{section:simulations}, migrating some parts of the codes to more efficient hardware is another possibility as well (e.g. GPUs). However, such solutions require development time. Since the allocation of CPU time can be generous in superclusters, scientists tend to use the brute force time-consuming (hence carbon-emitting) methods whenever feasible. Note however that human time is carbon-emitting as well, and one year development time in order to make codes more efficient might be not negligible in terms of GHG emissions.
Encouraging the use of more efficient languages than Python, a commonly used language among astrophysicists, could also substantially reduce the emissions (by more than an order of magnitude), as discussed in Ref.~\cite{Zwart20}.

One important action would be to educate the collaboration members about the carbon cost of simulations, and to give them incentives to weigh the cost/benefit of their simulation runs. This could be enforced by making members ask for clearance from the collaboration to run very CPU expensive tasks on collaboration managed resources, by giving scientific justification. 

\subsubsection{Electronic devices}

Finally, concerning electronic devices, the collaboration could add a recommendation to use devices longer in the GRAND Green Policy. As can be clearly seen in Table~\ref{tab:devices}, using a laptop 1.5 year longer could depreciate the emissions of $\sim 30\%$.

\subsection{Hardware}

By definition, the hardware equipment relies strongly on technologies and on their changes. By the 2030s, it is clear that the energy production and storage technologies will have changed. Solar panels and batteries will be likely smaller, more efficient, and less carbon-emitting. Recycling processes will also be different. One important challenge for the collaboration will thus be to stay alert to innovations concerning solar panels and batteries, and to be flexible in the design of the next generation equipment, so as to be able to integrate any improved device.

Before entering any hardware production phase, the collaboration should carefully balance the environmental cost of the materials used, their recyclability and establish a recycling plan. For example, stainless steel remains the favorite material, as it is durable, sustainable and its recycling rate in China is around 70\% \cite{RecyclingRate}. Lead-acid batteries have an important environmental impact, and lead is classified as the one of the top heavy metal pollutants in China. However, around 30\% of the primary lead production can be avoided with proper management of the old lead-acid batteries \cite{recyclingBatteries}: the implementation of a recycling plan will be crucial for GRAND.

Recycling can be either included in the contract signed with the owner of the experimental site, or the collaboration could directly establish a contract with recycling companies. The administrative details will be discussed within the collaboration. Another key point that the collaboration wishes to carefully monitor is the working conditions under which the equipment is produced. This ethical point will also be considered prior to the production phase.

The electronic parts of the antennas will be optimized and the power consumption of the data acquisition system will be largely reduced in the next stages of the project. This, combined with more efficient solar panels to be developed in the coming years implies that the hardware size and weight will be reduced. The triangular box harboring the electronic system will become smaller and lighter. As of today, it is difficult to give a reliable estimate of the reduction rate. Figure~\ref{fig:proj_all} shows that stainless steel is the main source of emissions for GRAND10k and GRAND200k. The fact that a considerable reduction of stainless steel emissions can also be considered in the future stages is promising.

Finally it is important to mention  that hardware transportation across the world has not been added to our projections. Although hardware transportation represents a minor share in the total emissions of the GRANDProto300 and GRAND200k phases, worldwide shipping could considerably boost the importance of this emission source. The weight of stainless steel could be reduced, as discussed above, which would also help to reduce the related transportation emissions. Hardware transportation will have to be carefully monitored, and the collaboration should discuss for example if production has to happen in China or could be done locally in the countries hosting the sub-arrays. When possible, low-carbon emitting transportation should be used for long distances across countries. 

\section{Conclusions and perspectives}\label{section:conclusion}

Large-scale astrophysics experiments are a collection of carbon emitting sources and practices, in particular through extensive travels, massive digital usage and hardware production and deployment. Scientists have a duty to set an example for climate actions.  
It is hence important to evaluate and discuss the carbon footprint of the experiments that are being projected, and endeavor to reduce it.

In this paper we have carried out a pioneering estimate of the global GHG emissions of a large-scale Astrophysics experiment: GRAND. The Giant Array for Neutrino Detection (GRAND) project aims at detecting ultra-high energy neutrinos with a 200,000 radio antenna array over 200,000\,km$^2$. The total array will be split into $\sim 20$ sub-arrays of 10,000\,antennas each, located at different sites worldwide.

We calculated the emissions related to three unavoidable sources: travel, digital technologies and hardware equipment. 
We find that these emission sources have a different impact depending on the stages of the experiment. Digital technologies, and to a lesser extent travel, prevail for the small-scale prototyping phase (GRANDProto300), whereas hardware equipment (material production and transportation) and data transfer/storage largely outweigh the other emission sources in the large-scale phase (GRAND200k). In the mid-scale phase (GRAND10k), the three sources contribute equally.  

We caution that the numbers brought in this study are order-of-magnitude estimates which serve to balance the choices and priorities of the GRAND collaboration to reduce its emissions. The estimates are subject to large uncertainties, and the projections presented are expected to vary widely, due to unforeseen experimental, technological and also human, societal changes. It is our hope that the scientific and technological progress that will take place in the next decade will reduce the emissions from the hardware equipment and the digital technologies of our project.

Still this study highlights the considerable carbon footprint of a a large-scale Astrophysics experiment. This should however be put in perspective, and compared with the emissions related to the production of everyday life items: the yearly emissions related to GRAND would amount to less than that of the manufacturing of 1000 cars.

The GRAND project being still in its prototyping stage, the study provides guidance to the future collaborative practices and instrumental design in order to reduce its carbon footprint.
There is room for improvement of the overall carbon footprint. Various types of actions may be implemented to mitigate it at all stages of the project deployment.

Travel emissions may be reduced by encouraging local collaborators to perform the on-site missions or by having international collaborators stay longer on the site of the experiment; they may also be reduced by optimizing collaboration meetings, through optimizing the location of the meetings, limiting the number of attendees from the collaboration, opting for some virtual meetings, and combining virtual and physical meetings. This discussion about collaboration meetings is similar to that being held by the Astrophysics community \cite{williamson2019embedding,matzner2019astronomy,stevens2019imperative,barret2020estimating,Labo1.5}

Options to reduce digital emissions include the reduction in the volume of data to be archived. 
The collaboration is already developing data reduction strategies to reduce the carbon footprint of data transfer and storage by 4 or 5 orders of magnitude. As for the emissions from simulations and data analysis, the challenge is to reduce the millions of CPU hours projected to be spent yearly. Incentives to weigh the cost/benefit of the simulations runs may contribute to lower the carbon footprint in the years to come.  

Mitigating the emissions from manufacturing and hauling the hardware equipment will be a top priority for the design of the GRAND200k phase, as these emissions are projected to weigh most in the carbon footprint of GRAND200k phase. It is about balancing the environmental cost of the materials used for the antennas, the solar panels and the batteries, establishing a recycling plan, and monitoring the transportation from the production sites to the array-sites.

The GRAND collaboration is planning to take several actions in response to this study. The various action plans proposed for each emission source will be documented in a GRAND Green Policy, which each collaboration member will be encouraged to follow, in order to reduce the collective carbon footprint. 

\section*{Acknowledgements}
We are grateful to Françoise Berthoud, Anna Risch, and to many GRAND collaborators (in particular Rafael Alves Batista, Aur\'elien Benoit-L\'evy, Simon Chiche, Valentin Decoene, Claire Gu\'epin, Eric Hivon, Olivier Martineau, Valentin Niess, Simon Prunet, Markus Roth, Charles Timmermans, Matias Tueros) for very fruitful discussions and ideas. This work was kindly supported by the direction of IAP and the GReCO team. This work is supported by the APACHE grant (ANR-16-CE31-0001) of the French Agence Nationale de la Recherche.

\appendix
\section{Electricity emission factors of various countries}\label{app:Emix}

We present in Table~\ref{tab:energymix} the different energy mixes for the countries that host GRAND collaboration members~\cite{ADEMECarbonDatabase}. Some of these numbers are used in this work for the calculation of the emissions from numerical simulations (Section~\ref{section:simulations}). France has a low-carbon electricity mix owing to its nuclear fleet, and Brazil thanks to hydropower and other renewables. In the other countries presented in this table, energy is primarily sourced by fossil fuel. 

\begin{table}
\centering
\begin{tabular}{lr}
  \hline
  Country & Electricity emission factor  \\
  & [kgCO$_{2}$e/kWh]\\
  \hline \hline
  Brazil & 0.0868  \\
  China & 0.766  \\
  France & 0.0571 \\
  Germany & 0.461 \\
  Netherlands & 0.415  \\
  United-States & 0.522  \\
  \hline

\end{tabular}
\caption{Electricity emission factors of different countries that host GRAND collaboration members. Source: ADEME Carbon Database (2020)~\cite{ADEMECarbonDatabase}.} 
  \label{tab:energymix}
\end{table}

\section{Global internet electricity emission factor estimate}\label{app:data_centers}

In Section~\ref{section:data_transfer}, we calculate the GHG emissions produced by data transfer via the Internet. The emission factor for data transmission is dependent on the energy mix used. As there is no data for the energy mix of the electricity used for the Internet, which is a global phenomenon, we chose to compute our own electricity emission factor using the distribution of data centers across the world. This distribution is documented in Ref.~\cite{datacenter}. We chose to focus on the 6 countries which host the majority of data centers in number (we did not take into account the size of these data centers). The number of data centers and their world share for these 6 countries are given in Table~\ref{tab:data_centers}. Along with the electricity emission factor of each of these countries, we used their weight (world share) to compute a general emission factor. The electricity emission factor of each country stems from the ADEME Carbon Database (2020)~\cite{ADEMECarbonDatabase}.

\begin{table}
\centering
\resizebox{\columnwidth}{!}{%
\begin{tabular}{lrrr}
  \hline
  Country & Electrity emission factor & Data center  & World  \\
  & [kgCO$_{2}$e/kWh] & number & share   \\
  \hline \hline
  USA & 0.522 & 1778 & 38\%  \\
  \hline
  UK & 0.457 & 272 & 3\%  \\
  \hline
  France & 0.0571 & 154 & 3\%  \\
  \hline
  India & 0.912 & 149 & 3\%  \\
  \hline
  Australia & 0.841 & 117 & 2\% \\
  \hline
  Netherlands & 0.415 & 113 & 2\%  \\
  \hline
  
\end{tabular}}
\caption{List of the 6 countries hosting the majority of data centers in the world (in number, not taking into account their size)  used to calculate a global electricity emission factor of the Internet. The number of data centers, their world share (number of data centers divided by the total number worldwide, in percentage), and the electricity emission factor corresponding to the country~\cite{ADEMECarbonDatabase} are indicated.} 
  \label{tab:data_centers}
\end{table}

\bibliography{biblio}

\begin{thebibliography}{10}
\expandafter\ifx\csname url\endcsname\relax
  \def\url#1{\texttt{#1}}\fi
\expandafter\ifx\csname urlprefix\endcsname\relax\def\urlprefix{URL }\fi
\expandafter\ifx\csname href\endcsname\relax
  \def\href#1#2{#2} \def\path#1{#1}\fi

\bibitem{williamson2019embedding}
K.~{Williamson}, T.~A. {Rector}, J.~{Lowenthal}, {Embedding Climate Change
  Engagement in Astronomy Education and Research}, in: Bulletin of the American
  Astronomical Society, Vol.~51, 2019, p.~49.
\newblock \href {http://arxiv.org/abs/1907.08043} {\path{arXiv:1907.08043}}.

\bibitem{matzner2019astronomy}
C.~{Matzner}, N.~B. {Cowan}, R.~{Doyon}, V.~{H{\'e}nault-Brunet},
  D.~{Lafreni{\`e}re}, M.~{Lokken}, P.~G. {Martin}, S.~{Morsink},
  M.~{Nomandeau}, N.~{Ouellette}, M.~{Rahman}, J.~{Roediger}, J.~{Taylor},
  R.~{Thacker}, M.~{van Kerkwijk}, {Astronomy in a Low-Carbon Future}, in:
  Canadian Long Range Plan for Astronomy and Astrophysics White Papers, Vol.
  2020, 2019, p.~22.
\newblock \href {http://arxiv.org/abs/1910.01272} {\path{arXiv:1910.01272}},
  \href {http://dx.doi.org/10.5281/zenodo.3758549}
  {\path{doi:10.5281/zenodo.3758549}}.

\bibitem{stevens2019imperative}
A.~R.~H. {Stevens}, S.~{Bellstedt}, P.~J. {Elahi}, M.~T. {Murphy}, {The
  imperative to reduce carbon emissions in astronomy}, Nature Astronomy 4
  (2020) 843--851.
\newblock \href {http://arxiv.org/abs/1912.05834} {\path{arXiv:1912.05834}},
  \href {http://dx.doi.org/10.1038/s41550-020-1169-1}
  {\path{doi:10.1038/s41550-020-1169-1}}.

\bibitem{barret2020estimating}
D.~{Barret}, {Estimating, monitoring and minimizing the travel footprint
  associated with the development of the Athena X-ray Integral Field Unit},
  Experimental Astronomy 49~(3) (2020) 183--216.
\newblock \href {http://arxiv.org/abs/2004.05603} {\path{arXiv:2004.05603}},
  \href {http://dx.doi.org/10.1007/s10686-020-09659-8}
  {\path{doi:10.1007/s10686-020-09659-8}}.

\bibitem{Labo1.5}
J.~Mariette, O.~Blanchard, O.~Bern{\'e}, T.~Ben-Ari,
  \href{https://labos1point5.org/}{{An open-source tool to assess the carbon
  footprint of research.}}
\newblock \href {http://arxiv.org/abs/2021.01.14.426384}
  {\path{arXiv:2021.01.14.426384}}, \href
  {http://dx.doi.org/10.1101/2021.01.14.426384}
  {\path{doi:10.1101/2021.01.14.426384}}.
\newline\urlprefix\url{https://labos1point5.org/}

\bibitem{GRAND20}
{GRAND Collaboration}, J.~{{\'A}lvarez-Mu{\~n}iz}, R.~{Alves Batista},
  A.~{Balagopal V.}, et~al., {The Giant Radio Array for Neutrino Detection
  (GRAND): Science and design}, Science China Physics, Mechanics, and Astronomy
  63~(1) (2020) 219501.
\newblock \href {http://arxiv.org/abs/1810.09994} {\path{arXiv:1810.09994}},
  \href {http://dx.doi.org/10.1007/s11433-018-9385-7}
  {\path{doi:10.1007/s11433-018-9385-7}}.

\bibitem{GHGprotocol}
\href{https://ghgprotocol.org/}{{Greenhouse gas protocol}}.
\newline\urlprefix\url{https://ghgprotocol.org/}

\bibitem{IPCC18}
V.~Masson-Delmotte, P.~Zhai, H.-O. P{\"o}rtner, D.~Roberts, J.~Skea, P.~Shukla,
  A.~Pirani, W.~Moufouma-Okia, C.~P{\'e}an, R.~Pidcock, S.~Connors,
  J.~Matthews, Y.~Che, X.~Zhou, M.~Gomis, E.~Lonnoy, T.~Maycock, M.~Tignor,
  T.~Waterfield, Summary for policymakers. in: Global warming of
  1.5$\,^{\circ}$c. an ipcc special report on the impacts of global warming of
  1.5$\,^{\circ}$c above pre-industrial levels and related global greenhouse
  gas emission pathways, in the context of strengthening the global response to
  the threat of climate change, sustainable development, and efforts to
  eradicate poverty, World Meteorological Organization (2018) 32.

\bibitem{IPCC5}
T.~Stocker, D.~Qin, G.-K. Plattner, M.~Tignor, S.~Allen, J.~Boschung,
  A.~Nauels, Y.~Xia, V.~Bex, P.~Midgley, {IPCC, 2013: Climate Change 2013: The
  Physical Science Basis. Contribution of Working Group I to the Fifth
  Assessment Report of the Intergovernmental Panel on Climate Change},
  Cambridge University Press (2013) 1535.

\bibitem{CO2eq}
C.~M. Brander,
  \href{https://ecometrica.com/assets/GHGs-CO2-CO2e-and-Carbon-What-Do-These-Mean-v2.1.pdf}{{Greenhouse
  Gases, CO2, CO2e, and Carbon:What Do All These Terms Mean?}} (2012).
\newline\urlprefix\url{https://ecometrica.com/assets/GHGs-CO2-CO2e-and-Carbon-What-Do-These-Mean-v2.1.pdf}

\bibitem{methodo}
{Association Bilan Carbone},
  \href{{https://www.associationbilancarbone.fr/wp-content/uploads/2018/03/bilan-carbone-v8-guide-methodologique-final.pdf}}{{Guide
  m{\'e}thodologique : Bilan Carbone V8, objectifs et principes de
  comptabilisation}} (2017).
\newline\urlprefix\url{{https://www.associationbilancarbone.fr/wp-content/uploads/2018/03/bilan-carbone-v8-guide-methodologique-final.pdf}}

\bibitem{IEA}
{International Energy Agency},
  \href{https://www.iea.org/reports/tracking-transport-2019}{{Tracking
  Transport 2018}} (2018).
\newline\urlprefix\url{https://www.iea.org/reports/tracking-transport-2019}

\bibitem{ADEMECarbonDatabase}
ADEME, \href{http://www.bilans-ges.ademe.fr}{Carbon data base}.
\newline\urlprefix\url{http://www.bilans-ges.ademe.fr}

\bibitem{IPCC1999}
J.~Penner, D.~Lister, D.~Griggs, D.~Dokken, M.~McFarland,
  \href{https://www.ipcc.ch/report/aviation-and-the-global-atmosphere-2/}{Aviation
  and the global atmosphere}, prepared in collaboration with the Scientific
  Assessment Panel to the Montreal Protocol on Substances that Deplete the
  Ozone Layer (1999).
\newline\urlprefix\url{https://www.ipcc.ch/report/aviation-and-the-global-atmosphere-2/}

\bibitem{MinistryEcology}
{French Ministry for an ecological and solidary Transition},
  \href{https://www.ecologique-solidaire.gouv.fr/sites/default/files/Information_GES\%20-\%202019.pdf}{Ghg
  information for transport services} (2019).
\newline\urlprefix\url{https://www.ecologique-solidaire.gouv.fr/sites/default/files/Information_GES\%20-\%202019.pdf}

\bibitem{KLM}
\href{{https://www.klm.com/travel/nl_en/prepare_for_travel/fly_co2_neutral/all_about_sustainable_travel/index.htm}}{Klm
  data} (2019).
\newline\urlprefix\url{{https://www.klm.com/travel/nl_en/prepare_for_travel/fly_co2_neutral/all_about_sustainable_travel/index.htm}}

\bibitem{MyClimate}
\href{https://www.myclimate.org/fileadmin/user_upload/myclimate_-_home/01_Information/01_About_myclimate/09_Calculation_principles/Documents/myclimate-flight-calculator-documentation_EN.pdf}{The
  myclimate flight emission calculator} (2019).
\newline\urlprefix\url{https://www.myclimate.org/fileadmin/user_upload/myclimate_-_home/01_Information/01_About_myclimate/09_Calculation_principles/Documents/myclimate-flight-calculator-documentation_EN.pdf}

\bibitem{ShiftProject}
{The Shift Project},
  \href{https://theshiftproject.org/en/article/lean-ict-our-new-report/}{{Lean
  ICT - Towards digital sobriety}} (2019).
\newline\urlprefix\url{https://theshiftproject.org/en/article/lean-ict-our-new-report/}

\bibitem{ecodiag}
EcoInfo, \href{{https://ecoinfo.cnrs.fr/ecodiag-calcul/}}{Ecodiag} (2020).
\newline\urlprefix\url{{https://ecoinfo.cnrs.fr/ecodiag-calcul/}}

\bibitem{video}
D.~Ong, T.~Moors, V.~Sivaraman, Comparison of the energy, carbon and time costs
  of videoconferencing and in-person meetings, Computer Communications 50.
\newblock \href {http://dx.doi.org/10.1016/j.comcom.2014.02.009}
  {\path{doi:10.1016/j.comcom.2014.02.009}}.

\bibitem{2020NatAs...4..823B}
L.~{Burtscher}, D.~{Barret}, A.~P. {Borkar}, V.~{Grinberg}, K.~{Jahnke},
  S.~{Kendrew}, G.~{Maffey}, M.~J. {McCaughrean}, {The carbon footprint of
  large astronomy meetings}, Nature Astronomy 4 (2020) 823--825.
\newblock \href {http://arxiv.org/abs/2009.11344} {\path{arXiv:2009.11344}},
  \href {http://dx.doi.org/10.1038/s41550-020-1207-z}
  {\path{doi:10.1038/s41550-020-1207-z}}.

\bibitem{2020Natur.583..356K}
M.~{Kl{\"o}wer}, D.~{Hopkins}, M.~{Allen}, J.~{Higham}, {An analysis of ways to
  decarbonize conference travel after COVID-19}, Nature 583~(7816) (2020)
  356--359.
\newblock \href {http://dx.doi.org/10.1038/d41586-020-02057-2}
  {\path{doi:10.1038/d41586-020-02057-2}}.

\bibitem{Berthoud}
F.~Berthoud, B.~Bzeznik, N.~Gibelin, M.~Laurens, C.~Bonamy, M.~Morel,
  X.~Schwindenhammer,
  \href{https://hal.archives-ouvertes.fr/hal-02549565}{{Estimation de
  l'empreinte carbone d'une heure.coeur de calcul}}.
\newline\urlprefix\url{https://hal.archives-ouvertes.fr/hal-02549565}

\bibitem{USA_Storage}
A.~Shehabi, S.~Smith, N.~Horner, I.~Azevedo, R.~Brown, J.~Koomey, E.~Masanet,
  D.~Sartor, M.~Herrlin, W.~Lintner,
  \href{https://www.osti.gov/servlets/purl/1372902}{United states data center
  energy usage report}, Lawrence Berkeley National Laboratory, Berkeley,
  California.
\newline\urlprefix\url{https://www.osti.gov/servlets/purl/1372902}

\bibitem{ubiquiti}
\href{https://www.ui.com/airmax/bullet-ac/}{{Ubiquiti Airfiber}}.
\newline\urlprefix\url{https://www.ui.com/airmax/bullet-ac/}

\bibitem{InternetAslan}
J.~Aslan, K.~Mayers, J.~G. Koomey, C.~France,
  \href{https://onlinelibrary.wiley.com/doi/pdf/10.1111/jiec.12630}{Electricity
  intensity of internet data transmission: Untangling the estimates}, Journal
  of Industrial Ecology 22~(4) (2018) 785--798.
\newline\urlprefix\url{https://onlinelibrary.wiley.com/doi/pdf/10.1111/jiec.12630}

\bibitem{cloud_storage}
L.~Posani, A.~Paccoia, M.~Moschettini, The carbon footprint of distributed
  cloud storage (2019).
\newblock \href {http://arxiv.org/abs/1803.06973} {\path{arXiv:1803.06973}}.

\bibitem{PCB}
\href{https://www.matrixelectronics.com/products/panasonic/megtron-6/}{{MEGTRON6,
  High Speed, Low Loss Multi-layer Materials }}.
\newline\urlprefix\url{https://www.matrixelectronics.com/products/panasonic/megtron-6/}

\bibitem{metal}
T.~Norgate, S.~Jahanshahi, W.~Rankin, Assessing the environmental impact of
  metal production processes, Journal of Cleaner Production 15~(8-9) (2007)
  838--848.

\bibitem{JOHNSON2008181}
J.~Johnson, B.~Reck, T.~Wang, T.~Graedel,
  \href{http://www.sciencedirect.com/science/article/pii/S0301421507003655}{The
  energy benefit of stainless steel recycling}, Energy Policy 36~(1) (2008) 181
  -- 192.
\newline\urlprefix\url{http://www.sciencedirect.com/science/article/pii/S0301421507003655}

\bibitem{ISSF2019}
{International Stainless Steel Forum},
  \href{https://www.worldstainless.org/Files/issf/non-image-files/PDF/ISSF_Stainless_Steel_and_CO2.pdf}{Stainless
  steel and co2: Facts and scientific observations}.
\newline\urlprefix\url{https://www.worldstainless.org/Files/issf/non-image-files/PDF/ISSF_Stainless_Steel_and_CO2.pdf}

\bibitem{pv}
A.~Louwen, W.~G. Van~Sark, A.~P. Faaij, R.~E. Schropp, Re-assessment of net
  energy production and greenhouse gas emissions avoidance after 40 years of
  photovoltaics development, Nature Communications 7~(1) (2016) 1--9.

\bibitem{Lead-acid}
E.~McKenna, M.~McManus, S.~Cooper, M.~Thomson,
  \href{http://www.sciencedirect.com/science/article/pii/S0306261912008094}{Economic
  and environmental impact of lead-acid batteries in grid-connected domestic pv
  systems}, Applied Energy 104 (2013) 239 -- 249.
\newline\urlprefix\url{http://www.sciencedirect.com/science/article/pii/S0306261912008094}

\bibitem{Sodium}
J.~Peters, D.~Buchholz, S.~Passerini, M.~Weil,
  \href{http://dx.doi.org/10.1039/C6EE00640J}{Life cycle assessment of
  sodium-ion batteries}, Energy Environ. Sci. 9 (2016) 1744--1751.
\newblock \href {http://dx.doi.org/10.1039/C6EE00640J}
  {\path{doi:10.1039/C6EE00640J}}.
\newline\urlprefix\url{http://dx.doi.org/10.1039/C6EE00640J}

\bibitem{Auger}
\href{https://www.auger.org/}{{The Pierre Auger Observatory}}.
\newline\urlprefix\url{https://www.auger.org/}

\bibitem{Zilles20_RM}
A.~{Zilles}, O.~{Martineau-Huynh}, K.~{Kotera}, M.~{Tueros}, K.~{de Vries},
  W.~{Carvalho}, V.~{Niess}, N.~{Renault-Tinacci}, V.~{Decoene}, {Radio
  Morphing: towards a fast computation of the radio signal from air showers},
  Astroparticle Physics 114 (2020) 10--21.
\newblock \href {http://dx.doi.org/10.1016/j.astropartphys.2019.06.001}
  {\path{doi:10.1016/j.astropartphys.2019.06.001}}.

\bibitem{tueros2020synthesis}
M.~Tueros, A.~Zilles, Synthesis of radio signals from extensive air showers
  using previously computed microscopic simulations (2020).
\newblock \href {http://arxiv.org/abs/2008.06454} {\path{arXiv:2008.06454}}.

\bibitem{Car_manufacturing_EC}
\href{https://ec.europa.eu/clima/sites/clima/files/transport/vehicles/docs/2020_study_main_report_en.pdf}{{Determining
  the environmental impacts of conventional and alternatively fuelled vehicles
  through LCA}}.
\newline\urlprefix\url{https://ec.europa.eu/clima/sites/clima/files/transport/vehicles/docs/2020_study_main_report_en.pdf}

\bibitem{Wynes19}
S.~Wynes, S.~D. Donner, S.~Tannason, N.~Nabors, Academic air travel has a
  limited influence on professional success, Journal of Cleaner Production 226
  (2019) 959 -- 967.
\newblock \href
  {http://dx.doi.org/https://doi.org/10.1016/j.jclepro.2019.04.109}
  {\path{doi:https://doi.org/10.1016/j.jclepro.2019.04.109}}.

\bibitem{Resultats1point5}
{Blanchard, M. and Bouchet-Valat, M. and Cartron, D. and Greffion, J. and Gros,
  J.},
  \href{https://labos1point5.org/les-enquetes/enquete1-resultat/}{{Premiers
  r{\'e}sultats de l'enqu{\^e}te ``Les personnels de la recherche face au
  changement climatique, Labos 1point5}} (2021).
\newline\urlprefix\url{https://labos1point5.org/les-enquetes/enquete1-resultat/}

\bibitem{ResponsibleAcademia}
J.~Stroud, K.~Feeley, Responsible academia: optimizing conference locations to
  minimize greenhouse gas emissions, Ecography 38 (2016) 402--404.

\bibitem{MozillaHub}
\href{https://hubs.mozilla.com/}{Mozilla hub}.
\newline\urlprefix\url{https://hubs.mozilla.com/}

\bibitem{Organize}
O.~Reshef, I.~Aharonovich, A.~Armani, S.~Gigan, R.~Grange, M.~Kats,
  R.~Sapienza, How to organize an online conference, Nature Reviews Materials
  5.
\newblock \href {http://dx.doi.org/10.1038/s41578-020-0194-0}
  {\path{doi:10.1038/s41578-020-0194-0}}.

\bibitem{Low-carbon}
A.~Abbott, Low-carbon, virtual science conference tries to recreate social
  buzz, Nature 577.
\newblock \href {http://dx.doi.org/https://doi.org/10.1038/d41586-019-03899-1}
  {\path{doi:https://doi.org/10.1038/d41586-019-03899-1}}.

\bibitem{Zwart20}
S.~Portegies~Zwart, \href{https://doi.org/10.1038/s41550-020-1208-y}{The
  ecological impact of high-performance computing in astrophysics}, Nature
  Astronomy 4~(9) (2020) 819--822.
\newblock \href {http://dx.doi.org/10.1038/s41550-020-1208-y}
  {\path{doi:10.1038/s41550-020-1208-y}}.
\newline\urlprefix\url{https://doi.org/10.1038/s41550-020-1208-y}

\bibitem{RecyclingRate}
B.~K. Reck, M.~Chambon, S.~Hashimoto, T.~Graedel,
  \href{https://doi.org/10.1021/es903584q}{Global stainless steel cycle
  exemplifies china's rise to metal dominance}, Environmental Science \&
  Technology 44~(10) (2010) 3940--3946, pMID: 20426460.
\newblock \href {http://arxiv.org/abs/https://doi.org/10.1021/es903584q}
  {\path{arXiv:https://doi.org/10.1021/es903584q}}, \href
  {http://dx.doi.org/10.1021/es903584q} {\path{doi:10.1021/es903584q}}.
\newline\urlprefix\url{https://doi.org/10.1021/es903584q}

\bibitem{recyclingBatteries}
Z.~Sun, H.~Cao, X.~Zhang, W.~Zheng, G.~Cao, Y.~Sun, Y.~Zhang, Spent lead-acid
  battery recycling in china -- a review and sustainable analyses on mass flow
  of lead, Waste Management 64.
\newblock \href {http://dx.doi.org/10.1016/j.wasman.2017.03.007}
  {\path{doi:10.1016/j.wasman.2017.03.007}}.

\bibitem{datacenter}
\href{https://www.datacentermap.com/datacenters.html}{{Colocation Data
  Centers}} (2020).
\newline\urlprefix\url{https://www.datacentermap.com/datacenters.html}

\end{thebibliography}
\end{document}